\documentstyle[aps,eqsecnum,epsf,floats,twocolumn]{revtex}

\begin{document}

\def\Spicture#1#2#3{\begin{tabular}{c}{\epsfxsize=#1\epsfbox{#2}}\\
\mbox{#3}\end{tabular}}

\def\leftline{\hfill
\begin{tabular}{|ccc}
\hline
$\ \ \ \ \ \ \ \ \ \ \ \ \ \ $  
\hfill 
$\ \ \ \ \ \ \ \ \ \ \ \ \ \ $
& 
$\ \ \ \ \ \ \ \ \ \ \ \ \ \ $ 
\hfill 
$\ \ \ \ \ \ \ \ \ \ \ \ \ \ $
& \hfill 
$\ \ \ \ \ \ \ \ \ \ \ \ \ \ $ \\
\end{tabular}
}

\def\leftlineup{\hfill
\begin{tabular}{|ccc}
$\ \ \ \ \ \ \ \ \ \ \ \ \ \ $  
\hfill 
$\ \ \ \ \ \ \ \ \ \ \ \ \ \ $
& 
$\ \ \ \ \ \ \ \ \ \ \ \ \ \ $ 
\hfill 
$\ \ \ \ \ \ \ \ \ \ \ \ \ \ $
& \hfill 
$\ \ \ \ \ \ \ \ \ \ \ \ \ \ $ \\
\hline
\end{tabular}
}

\def\rightline{\hfill
\begin{tabular}{ccc|}
$\ \ \ \ \ \ \ \ \ \ \ \ \ \ $  
\hfill 
$\ \ \ \ \ \ \ \ \ \ \ \ \ \ $
& 
$\ \ \ \ \ \ \ \ \ \ \ \ \ \ $ 
\hfill 
$\ \ \ \ \ \ \ \ \ \ \ \ \ \ $
& \hfill 
$\ \ \ \ \ \ \ \ \ \ \ \ \ \ $ \\
\hline
\end{tabular}
}

\draft
\tightenlines

\twocolumn[\hsize\textwidth\columnwidth\hsize\csname
@twocolumnfalse\endcsname

\preprint
{$\ ^{\tt DFPD\ 98/TH/22\ 
(University\ of\ Padova)}$}

\title{  {The pion-three-nucleon problem
with two-cluster connected-kernel equations
 }}

 \author{  L. Canton  }
 \address{  Istituto Nazionale di Fisica Nucleare\\
      via Marzolo 8, Padova I-35131, Italia  }

\date{June, 1$^{st}$ 1998}

\maketitle

\widetext

 \begin{abstract}
It is found that the coupled $\pi$NNN-NNN system breaks into fragments 
in a nontrivial way. Assuming the particles as distinguishable,
there are indeed four modes of fragmentation into two clusters,  
while in the standard three-body problem there are three possible 
two-cluster partitions and conversely the four-body problem has 
seven different possibilities.
It is shown how to formulate
the pion-three-nucleon collision problem through the 
integral-equation approach by taking into account the proper fragmentation
of the system. 
The final result does not depend on the assumption of separability of the 
two-body t-matrices.
Then, the quasiparticle method {\em \`a la} Grassberger-Sandhas is applied
and effective two-cluster connected-kernel equations are obtained.
The corresponding bound-state problem is also formulated,
and the resulting homogeneous equation provides a new approach
which generalizes the commonly used techniques
%(with or without three-body forces) 
to describe the three-nucleon bound-state 
problem, where the meson degrees of freedom are usually suppressed. 
 \end{abstract}


 %\newpage

]

\narrowtext

 \section{  Introduction  }

 \label{intro}

In the past years, there have been various attempts
to generalize the integral-equation approach to the
quantum few-body problem, and specifically the N-body formulation
of Sandhas and collaborators~\cite{AGS,GS}, to obtain
a formulation of the pion-three-nucleon problem
with the aim to handle this different problem 
(where the number of particles is not fixed)
with the same non-perturbative computational techniques
which have been developed and widely tested in standard 
few-body applications.

In the standard N-body approach, as is well known,  repeated 
applications of Faddeev's three-body treatment~\cite{faddeev} 
and, in every step,
of the quasiparticle method~\cite{sandhas}, lead to 
effective two-body equations for the collision processes
between composite particles. 
Few authors~\cite{AM,ueda} some years ago proposed a treatment of the
$\pi$NNN problem where quasiparticle equations
were assumed from the very beginning as a starting ansatz.
The treatment of Ref.\cite{AM}
started from the coupled $\pi$NNN-NNN dynamics and successfully 
arrived at the
first connected-kernel integral formulation of the problem, however
two-body equations describing binary 
collisions between composite particles of the {\em complete } system 
have not been obtained, since
the amplitudes were represented in terms of cluster
partitions of the four- and three-body spaces as if these were
two completely disjoint sectors.
In Ref.\cite{ueda}
the underlying three- and four-body dynamics
has been approximated by phenomenological multi-cluster two- and three-body
relativistic equations, including a 24-channel
effective two-body equation which then was  
solved numerically and compared with pion production data; 
however in this case 
it was not possible to show that the
approach is linked to or can be directly obtained from 
the underlying three- and four-body dynamics.

More recently, there has been another attempt to find
a better formulation of the coupled $\pi$NNN-NNN
problem~\cite{cc94ab}. The approach is more general than
the previous ones since it does not assume from the beginning
the quasiparticle (separable) ansatz but relies 
on the four-body chain-labelled formalism {\em \`a la} 
Yakubovsk\u{\i}~\cite{Y} and extends this formalism to the $\pi$NNN situation
where the pion can disappear through the $\pi$NN vertex interaction.
In this case, by repeated use of the quasiparticle method
(in close analogy with the standard four-body formulation~\cite{GS})
effective two-body equations for the collision problem 
between composite fragments of the whole system have been found,
but it has been shown in a subsequent analysis~\cite{cc97} that
{\em (i)} the leading equation has a disconnected kernel
and {\em (ii)} the amplitudes referring to the various 
rearrangement processes have intrinsic ambiguities and cannot be univocally
identified with the physical collision processes.
Both problems cannot be solved in that formalism
unless one disregards certain diagrams
referring to the 2+2 partitions, thereby making an approximation
which at the least breaks unitarity. 

Since all the above mentioned approaches achieved
only a limited success in the attempt to generalize
the Grassberger-Sandhas transition operator formalism
(or the equivalent Faddeev-Yakubovsk\u{\i} Green's function
formalism) to the pion-three-nucleon problem, 
one may arise the question whether
these multiparticle approaches are well suited to treat
the multinucleon dynamics in presence of an absorbable pion.
This paper is mainly focused on this important question
and arrives at an affirmative (although not general) conclusion:
It is indeed possible to generalize the 
Faddeev-Yakubovsk\u{\i}-Alt-Grassberger-Sandhas  
formalism, developed for the quantum-mechanical 
treatment of a fixed number of bodies, to the case
of the pion-multinucleon dynamics, at least
under the assumption that the proper Fock space
with its infinite number of particles (unavoidable
whenever production and/or absorption occurs)
is truncated and the sole states with at most one dynamical
pion are retained.
The formalism illustrated in the next section is indeed
an approximate, effective description of the three-nucleon 
collisional problem below the production threshold
of the second pion, and within the limits
set by the truncation of the Hilbert space to three and four particles
it is shown that it is possible to 
%generalize the Grassberger-Sandhas formalism and 
obtain the formal solution of the coupled 
$\pi$NNN-NNN collisional problem in terms of effective two-cluster 
connected-kernel equations. 

The approach begins with a set of equations, originally developed for the 
coupled $\pi$NN-NN problem by Thomas and Rinat~\cite{tr79},
and later extended by Afnan and Blankleider~\cite{AB},
and following a somewhat different method by
Avishai and Mizutani~\cite{am79}.
Their final equations merge the three-body dynamics in the $\pi$NN sector
together with the two-body dynamics of the NN sector and provide the
additional couplings between  the two sectors.
The introduction of the quasiparticle (separable) ansatz
for the two-body t-matrices allows to derive two-body effective equations
coupling together the two-cluster partitions of the whole system, 
in close similarity with the AGS~\cite{AGS} quasiparticle formalism
for the pure three body problem.

An important aspect of this formalism is that it satisfies
unitarity by construction
at both two- and three-body level~\cite{AB,am79}
provided that for the input two-body t-matrices
the off-shell unitarity relation is assumed,
that the Green's functions in the no pion sector
include the pion-loop self-energy diagrams,
that the $\pi$NN vertex is properly dressed with the
contribution coming from the non-polar $\pi$N interaction
and that at least the OPE contribution of the NN interaction
is treated nonstatically.
All these features have been carefully maintained~\cite{ccs} 
in the equations herein used as input
for the pion-three-nucleon problem.

Another aspect worth to mention here is that the relativistic 
dynamics of the system can be incorporated in these sets of equations
by modification of the Green's functions, along the lines
of the relativistic three-particle isobar approach in the 
Aaron-Amado-Young model, or by using the Blankenbecker-Sugar
reduction method to eliminate the time component from the 
integration variables in the four-dimensional covariant
equations. We refer to the books~\cite{garcimizu,adikowa}
and to the references contained therein for these possible
relativistic reformulations of the problem.

It must be acknowledged, on the other hand, that
in spite of all these attractive features
the input equations we start with are not
free from conceptual problems. One is connected to the unavoidable 
truncation of a time-ordered field theory to a limited
number of particles and is known as the nucleon renormalization
problem~\cite{sauer,blank}. The problem has practical consequences
in that the effective $\pi$NN coupling constant in the multinucleon 
media becomes systematically smaller than the one used as input 
to describe the pion-nucleon subsystem dynamics.
There are methods~\cite{kb,phianf} to handle this 
difficult problem but they will not be discussed here.

In Sect.~\ref{clusdec} the four partitions of the whole 
system into two clusters are introduced for the first time.
This partition mode has no counterparts either in the four-body 
sector (where there are seven two-cluster partitions)
or in the three-body sector (three two-cluster partitions)
but allows the two sectors to dialogue.

Then, to obtain the new integral-equation formulation, 
the following steps are taken:
Firstly, {\em } the input equations are reformulated in
a matrix Lippmann-Schwinger-type (LS) form where the role of the t-matrix,
(denoted ${\bf T}^{(3)}$ in matrix notation)   
is played by the multiparticle transition amplitudes
referring to all possible three-cluster partitions of the system. 
Secondly, {\em } the dynamical equations (again in LS form)
for the subsystems identified by two-cluster partitions 
are introduced.
Then {\em } a new sum rule is introduced with respect to the two-cluster
partition index, for the ``generalized" potential ${\bf V}^{(3)}$ (that is,
the operator that plays the role of the potential in the input LS equation).
Subsequently, {\em } from the set of three-cluster amplitudes ${\bf T}^{(3)}$ 
the two-cluster disconnected contributions are extracted.
Finally, by means of the previous results, {\em } a new equation for the 
remaining connected part of ${\bf T}^{(3)}$ has been derived.
The result by no means relies on the assumptions that the subsytem
t-matrices or amplitudes have a separable structure and therefore
it holds in general.

It is to be noted that in the standard N-body problem 
it is possible to recover the whole GS
multiparticle formulation, and 
rederive their final connected-kernel equations by recursive 
application of the procedure made by the steps just mentioned above,
since this recursive procedure allows to extract from the N-body
collision amplitude the whole set of disconnected contributions
ranging from the highest level (corresponding to partitions of the 
system in N-1 clusters), up to the lowest level of disconnectedness 
where the system is partitioned into two clusters~\cite{cc94ab}.
This fact emphasizes the close analogies between the GS formulation
and the approach here adopted to solve the $\pi$NNN problem.

In Sect.~\ref{quasiparticle} the quasiparticle
formalism is introduced. 
The quasiparticle method is applied once in the four-body sector 
and a second time simultaneously in both three and four-body sectors,
to exhibit diagrammatically the connected-kernel structure of the theory,
and to recast the result in terms of coupled multiparticle equations
for the two-cluster dynamics, since this is physically more transparent
and easier to communicate.
The equations are discussed in terms of coalescence
diagrams and particular attention is payed to the nonstandard
role of the pion. All the driving terms of the final two-cluster
coupled equations are exchange-type diagrams 
and are shown to connect the entire set of equations.
 
In Sect.~\ref{boundsys} the bound-state equation
for the coupled $\pi$NNN-NNN system is derived. As is well known, in the 
two-nucleon system the bound-state wavefunction can be expressed
as the negative-energy solution of the homogeneous equation whose kernel is
transposed with respect to that of the two-body LS equation, and similarly
the three-nucleon bound-state wavefunction can be expressed
in terms of the negative-energy solution of the homogeneous
equation whose kernel is transposed with respect to that of the AGS
equation. The homogeneous solution of the coupled $\pi$NN-NN
equations provides the natural way to include the pion dynamics
in the two-nucleon bound-state wavefunction, and from this fact it is shown
that it is possible to derive a three-nucleon bound-state
wavefunction (explicitly including the pion dynamics) which can be given
as solution of a new homogeneous equation whose kernel is 
similarly related to that of the equation we have derived in 
Sect.~\ref{clusdec} for the multiparticle collision problem.
If we switch off the couplings due to the $\pi$NN vertices
the homogeneous equation splits into two independent ones 
(with of course two independent spectra): one whose kernel is 
referable to the Faddeev-AGS one for the pure three-nucleon sector, 
and another homogeneous Yakubowsk\u{\i}-GS-type equation 
for the pure four-particle bound state. With the complete equation
it is possible to merge together the three and four-particle
aspects of the problem, thus providing, for the three-nucleon system,
a bound-state equation of new structure
which generalizes the ones investigated so far.

The approach may also serve as guidance to develop a consistent formulation
which divide the interaction of the three-nucleon system between a two-
and a three-body force. In fact, the need of three-body forces
naturally arises in theories where the meson
degrees of freedom are suppressed and the three nucleons are depicted
as point-like quantum particles interacting via local two-body potentials.
The common procedure uses some symmetry principles
to evaluate certain three-nucleon irreducible diagrams, selected
on physical grounds to give the dominant contribution to the 
three-body force
($\pi$N s-wave interaction at threshold~\cite{murcoon,afnasai}, or 
p-wave $\Delta$ excitation at intermediate energies~\cite{sau1,sau2}).
The approach here discussed performs
the complete resummation of the whole multiple scattering series
including all one-pion intermediate states, provides the 
source of all reducible and
irreducible three-nucleon  contributions of the one-pion type, 
and furthermore sets the proper framework for their 
nonperturbative handling.

In Sect.~\ref{amplitudes} the attention is put back to the collision 
problem and in particular to the rules for calculating the 
scattering amplitudes for all possible combinations of multiparticle
fragmentation involved in the collision.
Finally, in Sect.~\ref{conclusions} a brief summary and the conclusions
are given.

\section{  Cluster decomposition of the pion-three-nucleon system  }

 \label{clusdec}

 We consider as starting point the result obtained
    in Ref.\cite{ccs}.
 Here the dynamical equations coupling all the partitions
    of the $\pi$NNN system into three clusters
    have been derived following the diagrammatic approach
    and applying nontrivial properties of the four-body
    transition operators defined within the standard
    AGS theory.
    In this manner, it was possible to obtain
    an equation for new amplitudes where scattering processes, 
    pion production
    and absorption are coupled in an unitary treatment.

    The final coupled equations were formally identical to the
    Afnan-Blankleider (AB) equations, originally designed for the coupled
    $\pi$NN system:

\begin{mathletters}
\begin{eqnarray}
 U_{  ab  }&=&G_0^{  -1  }\bar\delta_{  ab  }+\sum_c\bar\delta_{  ac  }t_cG_0U_{  cb  }
 +F_a g_0 U^\dagger_b,
    \label{AB1} \\
    U^\dagger_a&=&F^\dagger_a+{  \cal V  }g_0U^\dagger_a+\sum_c F^\dagger_c
    G_0t_cG_0U_{  ca  },
    \label{AB2} \\
    U_a&=&F_a+\sum_c\bar\delta_{  ac  }t_cG_0U_c+F_ag_0U,
    \label{AB3}    \\
    U&=&{  \cal V  }+{  \cal V  }g_0U+\sum_c{  F^\dagger_c  }G_0t_cG_0U_c .
    \label{AB4}
   \end{eqnarray}
    \label{AB0}
   \end{mathletters}
%%%Bisogna spiegare il significato dei simboli
We briefly recall the meaning of the symbols, referring to Ref.~\cite{ccs}
and to the references therein contained for more detailed explanations.
The transition matrices $U_{ab}$ and $U$ represent the scattering amplitudes 
for the three-fragment collision processes in the four-particle
and three-nucleon sectors, respectively, while $U^\dagger_{a}$ and $U_{b}$
are the corresponding absorption and production amplitudes.

The two-body t-matrices acting between all the possible pairs 
(labelled ``$a$") of the four-particle sector are denoted by $t_a$,
while $F_a$ ($F^\dagger_a$) are calculated from the
elementary $\pi$NN production (absorption) vertices 
in a manner that is detailed below. As for the notation,
it must be observed that the absorption amplitude
$U_a^\dagger$ is not directly associated 
to the corresponding production amplitude via hermitean conjugation,
since the effect of complex conjugation on the boundary conditions
must be taken into account. The same considerations apply for the $\pi$NN
vertices, as these include the energy-dependent distortion effects
due to the non-polar $\pi$N interaction~\cite{AB}. 
Moreover we omit for conciseness the dependence 
upon the total energy of the system, $E$, since its role can be 
easily recovered by resorting to the analogy with the standard
few-body case.

The operator
$G_0$ represents the free four-body Green's function and
$g_0$ denotes the free three-nucleon Green's function
(with the inclusion of the pion self-energy contributions).
The boundary conditions are fixed by approaching the right-hand
cut in the complex energy plane from the above.  
Finally, ${ \cal V}$ represents the total interaction
acting amongst the three nucleons, and is given by the sum
over the three pairwise nuclear interactions, which must
include the nonstatic OPE diagrams.
For the sake of simplicity, we will not assume
the occurrence of a residual three-body force, although
irreducible three-nucleon forces can be - and indeed have 
already been - accommodated in formalisms of this sort \cite{AM};
we will however add in Sect.~\ref{conclusions}
a discussion on the subject under a general perspective.  
Eqs.~(\ref{AB0}) %,~\ref{AB2},~\ref{AB3}, and~\ref{AB4} 
can be viewed (or reinterpreted) as
    a generalized Lippmann-Schwinger equation;
 in fact if we restrict the description to the zero-pion
    sector, which corresponds to freezing the pion degrees of freedom,
    the set of equations collapses to the well known Lippmann--Schwinger equation
    describing the standard quantum--mechanical situation of nucleons
    interacting through the nuclear potential, {\em i.e.}
    \begin{equation}
 U={  \cal V  }+{  \cal V  }g_0U ,
    \end{equation}
and for the simpler two-nucleon system,  
$U$ corresponds to the well-known 
nucleon-nucleon t-matrix. Eqs.~(\ref{AB0})
%,~\ref{AB2},~\ref{AB3}, and~\ref{AB4} 
generalize the above equation by providing a direct
    link between the three-nucleon space and the three-cluster 
rearrangement processes in the four-particle space.
As is obvious, the index $a$ (or $b$, etc.) denotes the particle pair,
either $\pi N$ of NN, which form the composite fragment in the
four-body space. With $\bar\delta_{ab}(\equiv 1-\delta_{ab})$ it is 
meant 1 if the pairs $a$, $b$ are different, 0 otherwise.
    The link between the two spaces is made possible by the operators
    $F_a$ and $F_a^\dagger$, defined in terms of the elementary pion 
production or absorption vertices,
    \begin{equation}
 F_a=\sum_{  i=1  }^{  3  }\bar\delta_{  ia  } f_i,\ \ \ \ \
    F^\dagger_a=\sum_{  i=1  }^{  3  }\bar\delta_{  ia  } f^\dagger_i.
    \label{vertices}\end{equation}
Here, ``$i$" has a twofold meaning since it denotes the nucleon
which emits (or absorbs) the pion and at the same time the corresponding 
pion-nucleon pair. As mentioned above, the 
employed elementary vertices have to be dressed by the distortion 
effects of the non-polar contribution to the $\pi$N t-matrix,
$f_i=(1+t_iG_0) f^{(o)}_i$, and similar distortions apply for $f_i^\dagger$.

The analogy with the standard LS equation can be best exploited
   by formally rewriting the AB equations as a matrix LS
   equation
   \begin{equation}
{\bf  T  }^{  (3)  } ={\bf  V  }^{  (3)  } +
{\bf  V  }^{  (3)  } {\bf G  }_{  0  }^{  (3)  } {\bf  T  }^{  (3)  } ,
   \label{(N-1)-LS}
\end{equation}
where all operators are now 7$\times$7 matrix operators
   with indices spanning all the three--cluster partitions of the $\pi$NNN-NNN
    system. 
%    (see Tab.~\ref{tab1}). 
    This can be obtained by introducing the
   following definitions:
 \begin{eqnarray}
{  {\bf  G  }_{  0  }}^{  (3)  }
   \equiv&
      %\begin{array}{cc}
%{  G_{  0  }} ^{  (3)  }_{  (a|b)  } & {  G_{  0  }} ^{  (3)  }_{  (a|0)  } \cr
      % {  G_{  0  }} ^{  (3)  }_{  (0|b)  } & {  G_{  0  }} ^{  (3)  }_{  (0|0)  } \cr
      %\end{  array  } \right|
      %&=
      \left( \begin{array}{ccc  }
    G_0t_aG_0\delta_{  ab  } &\ & 0 \\
    0 &\ & g_0 \\
     \end{array} \right),
     \label{prop}\\
   {  {\bf  V  }}^{  (3)  }
   \equiv &
 %\left|
 %\begin{array}{cc}
   %{  V  } ^{  (3)  }_{  (a|b)  } & {  V  } ^{  (3)  }_{  (a|0)  } \cr
   %{  V  } ^{  (3)  }_{  (0|b)  } & {  V  } ^{  (3)  }_{  (0|0)  } \cr
   %\end{array} \right|
   %&=
   \left( \begin{array}{ccc}
  {  G_0  }^{  -1  }\bar\delta_{  ab  } &\ & F_a\\
  F^\dagger_b &\ &{  \cal V  } \\
  \end{array} \right),
  \\
  {  {\bf  T  }}^{  (3)  }
   \equiv
 &
  %\left|
 %\begin{array}{cc}
 %{  T  } ^{  (3)  }_{  (a|b)  } & {  T  } ^{  (3)  }_{  (a|0)  } \cr
 %{  T  } ^{  (3)  }_{  (0|b)  } & {  T  } ^{  (3)  }_{  (0|0)  } \cr
 %\end{array} \right|
 %&=
 \left( \begin{array}{ccc}
 U_{  ab  } &\ & U_a\\
U^\dagger_b &\ & U \\
\end{array} \right).
\end{eqnarray}

While for the $\pi$NN problem
the above equation is already connected and couples all the possible
two-cluster partitions
of the system (which include the two-nucleon state without pions), in the
$\pi$NNN case the same equation couples only 
three-cluster partitions, thus leading to the non 
connectedness of the equation. 
This problem can be immediately understood by reasoning in terms of classes of 
``disconnected" diagrams. In Eqs.~(\ref{AB0}) %,AB2,AB3,AB4}
all diagrams connecting only two of the four particles 
have been subtracted, via the t-matrices.  
These same diagrams, if considered in the $\pi$NN case, group the 
system into two fragments, hence all the remaining diagrams
contained in Eqs.~(\ref{AB0}) 
must connect the whole equation. However in the $\pi$NNN case 
such two-body diagrams arrange the system into three clusters; therefore
Eqs.~(\ref{AB0}) contain either diagrams
connecting the entire system, or diagrams 
arranging the system into two fragments. One has to isolate this last
class of diagrams of higher connectivity but still ``disconnected"
before the correct equation can be found.
This scenario is perfectly analogous
to the situation for the standard few-body problem, 
where the Faddeev-AGS equation
solves the three-body problem but leaves the four-body problem still out 
of reach. In the four-body problem one must introduce the partitions into
two clusters and repeat the same logical scheme
to obtain four-body connected-kernel equations of Yakubovsk\u{\i}-GS type.

From the above considerations it is clear that  
a great attention must be put first in finding the correct two-cluster
partitions for the system and then one can proceed toward
$\pi$NNN-NNN  connected-kernel equations.
Conversely, in the approach attempted previously
\cite{cc94ab} the two-cluster partitions
are identified literally with the seven two-cluster partitions 
of the standard four-body problem, while 
in the three-nucleon space  the homologous partitions were  
playing a secondary role.
That fragmentation scheme, depicted
in Tab.~\ref{tab2}, leads to the difficulties observed in 
Ref.~\cite{cc97}, where it was found that the resulting 
two-cluster amplitudes had intrinsic ambiguities  
and the kernel of the resolving equation
was not connected. Both aspects
originate from the same problem; the (non) proper
identification of the physical partitions of the complete 
system into two clusters. 

\begin{table}

\begin{center}

\begin{tabular}{|c|cc|}
\hline
$a'$ & $\pi$NNN sector & NNN sector \\
\hline
1 & $ N_1\ (N_2\ N_3\ \pi) $ & $N_1 (N_2\ N_3)$\\
2 & $ N_2\ (N_3\ N_1\ \pi) $ & $N_2 (N_3\ N_1)$\\
3 & $ N_3\ (N_1\ N_2\ \pi) $ & $N_3 (N_1\ N_2)$\\
4 & $(\pi\ N_1)\ (N_2\ N_3) $ & $N_1 (N_2\ N_3)$\\
5 & $(\pi\ N_2)\ (N_3\ N_1) $ & $N_2 (N_3\ N_1)$\\
6 & $(\pi\ N_3)\ (N_1\ N_2) $ & $N_3 (N_1\ N_2)$\\
7 & $\pi\ (N_3\ N_1\ N_2)   $ & $       -       $\\
\hline
\end{tabular}

\end{center}

\caption[*]{The seven two-cluster partitions of the $\pi$NNN-NNN system
in previous approaches.}

\label{tab2}

\end{table}

In the current approach, we identify only 4 two-cluster 
partitions, enlisted in Tab.~\ref{tab3}.
We label these partitions with the index $s$, spanning from
0 to 3. The partition $s$=0 represents the only genuine four-body 
partition of the $\pi$NNN system and corresponds to the last
partition reported in Tab.~\ref{tab2}. Here the pion is isolated
from the rest of the system, 
hence there is no direct coupling with the zero-pion sector.
The remaining partitions with $s$=1, 2, and 3 exhibit a
new structure with no counterparts in the standard few-body
theories. Each partition
represents a physical cluster decomposition which 
can be detected as an asymptotic channel and where, according to
Tab.~\ref{tab3},
one two-cluster no-pion state is coupled with 
two different two-cluster one-pion states.

\begin{table}

\begin{center}

\begin{tabular}{|c|cc|}
\hline
$s$ & $\pi$NNN sector & NNN sector \\
\hline
0 & $\pi\ (N_3\ N_1\ N_2)   $ & $       -       $\\
1 & $ N_1\ (N_2\ N_3\ \pi) $; $(\pi\ N_1)\ (N_2\ N_3)$  & $N_1 (N_2\ N_3)$\\
2 & $ N_2\ (N_3\ N_1\ \pi) $; $(\pi\ N_2)\ (N_3\ N_1)$  & $N_2 (N_3\ N_1)$\\
3 & $ N_3\ (N_1\ N_2\ \pi) $; $(\pi\ N_3)\ (N_1\ N_2)$  & $N_3 (N_1\ N_2)$\\
\hline
\end{tabular}
\end{center}
\caption[*]{The two-cluster partitions of the $\pi$NNN-NNN system
defined in this approach.}
\label{tab3}
\end{table}

%0
%Therefore the $\pi$NNN system breaks into two clusters in a  
%nontrivial way. It gives rise to 4 different partition modes,
%while the standard three--body problem has three possible two--cluster
%partitions and the four--body problem has seven possibilities.
%
%Once we have defined the modes of partition into subsystems,
We can now introduce the equations for the channel (or subsystem)
dynamics.
First we have to define the channel  interaction
$v_s$. (We will assume $s\ne 0$ since the $s=0$ case 
will be discussed separately with standard few-body techniques.)
When $s\ne 0$ the subsystem interaction couples the zero-pion sector
with the one-pion sector and one has to define the action of $v_s$
in each sector. In the one--pion sector $v_s$ is labelled
by the chain-of-partition index, $\{a' a\}$, where
$a'$ represents one of the possible partitions (two, for a given $s\ne 0$)
into two clusters of the four--body sector,
while $a$ represents one of the possible three--cluster partitions
which can be obtained from the sequential break-up of the partition $a'$.
Therefore, the structure of $v_s$ in the one-pion sector
can be best represented as
\begin{equation}
v_s=\left(v_s\right)_{a'a,b'b} ,
\end{equation}
where the partition indices fulfil the chain conditions 
$a\subset a'\subset s$ and $b\subset b'\subset s$.  
In the no-pion sector, the index $s$ is sufficient
to identify the two-cluster partition of the system,
since for $s\ne 0$ there is a one to one correspondence between
the index $s$ and the spectator nucleon, as can be directly
inferred from Tab.~\ref{tab3}. Thus, in the three-nucleon sector
we denote the two-nucleon potential by 
\begin{equation}
v_s=\left(v_s\right)_{-,-}.
\end{equation}
Up to now we have identified the diagonal blocks
of the channel interaction; however it is obvious that
the index structure of the diagonal block fixes unavoidably 
the structure of the off-diagonal couplings 
between the two sectors, {\em e.g.}
\begin{equation}
v_s=\left(v_s\right)_{a'a,-}.
\end{equation}

The way the channel interaction operates is rather remarkable and
deserves further comments:
note that if we drop all the explicit links to the one--pion sector
the interaction operator collapses to the standard two-nucleon interaction.
In this case, the one-pion sector affects the channel interaction 
only through the OPE diagram, being this  
explicitly included in the interaction.  
Thus, the present approach implies a highly nontrivial generalization of
what we identify as the NN potential in the three-nucleon system.
For a given $s\ne 0$, the nucleon-nucleon potential becomes a matrix 
operator acting not only as a standard two-nucleon potential
in the three-nucleon space, 
but acquires extra components and couplings to the chain-of-partition 
space of the four-body sector. For instance, for $s=1$, $v_s$
not only represents the standard NN potential between
nucleons 2 and 3, but has further couplings in the one--pion sector
to all possible sequential break-ups of the four--body system
which are allowed by the given $s$. And there is more than this.
In addition there is a fourth interaction term
(for $s=0$) which has no direct action in the three--nucleon
space since  it operates only in the four-body sector and
in particular in the chains of partitions obtained from the sequential 
break-up of the $\pi$ +  (NNN) channel.

Up to now we have discussed the general structure of the channel 
interactions,  but we have not given yet its explicit expressions.
To accomplish this we write
\begin{equation}
\left(v_s\right)_{a'a,b'b} = {G_0}^{-1}\bar\delta_{ab}\delta_{a'b'} 
	\delta_{a,b\subset a'}\delta_{a',b'\subset s}
\end{equation}
for the interaction in the one--pion sector, while
in the no-pion sector (only for $s\ne 0$) 
\begin{equation}
\left(v_s\right)_{-,-}  = {\cal V}_s \, ,
\end{equation}
denotes the pair potential  between the two 
interacting nucleons, representing the
nonstatic OPE diagram (as well as other possible static contributions 
which phenomenologically take into account more complicated diagrams
such as heavy-boson exchanges and/or multipion exchanges).
Finally, the off-diagonal interactions connecting the three-nucleon 
and four-body sectors are defined by
\begin{mathletters}
\begin{equation}
\left(v_s\right)_{a'a,-} = \sum_{i=1}^3 f_i \bar\delta_{ia}
	\delta_{i,a\subset a'}\delta_{a'\subset s}  \equiv
\left(f_s\right)_{a'a}
\end{equation}
and  
\begin{equation}
\left(v_s\right)_{-,b'b} = \sum_{i=1}^3 f^\dagger_i
        \bar\delta_{ib}
	\delta_{i,b\subset b'}\delta_{b'\subset s}\equiv
\left(f_s^\dagger\right)_{b'b} .
\end{equation}
\end{mathletters}
It must be observed that Tab.~\ref{tab3} is crucial
for discussing the structure of the subamplitudes.
For each $s\ne 0$, there are two two-cluster partitions in the 
four-body sector and one two-cluster partition in the three-nucleon sector.
Then, in the four-body sector, there are five possible sequential
break-up for a given $s$ (three when the partition is of type
3+1, and two when it is of the form 2+2), and in the three--nucleon sector
there is an additional one associated with the break-up of the
nucleonic pair. In conclusion we have a total amount of six components 
for each channel interaction with $s\ne 0$.
The case $s=0$ is obviously more simple, since the corresponding fragmentation
mode passes through one single two-cluster
 
%\leftline
\twocolumn[\hsize\textwidth\columnwidth\hsize\csname
@twocolumnfalse\endcsname
partition (of type 3+1)
of the four-body sector with no couplings to the three--nucleon sector. 
As is well known, this standard four-body partition has three
possible ulterior fragmentations into three clusters.
 The subsystem interaction $v_s$ for $s=0$  
couples together only these three components.

For each of these four different modes of fragmentation
into two clusters, we can introduce the subamplitudes, 
${\bf t}_s$, having the same 
chain-labelled structure of the channel interactions, 
with six components 
\begin{equation}
{\bf t}_s=\left(\begin{array}{ll}
(t_s)_{a'a,b'b}& (t_s)_{a'a,-}\\
	    (t_s)_{-,b'b}  & (t_s)_{-,-}
\end{array}\right)  
\equiv \left(
\begin{array}{ll}
(u_s)_{a'a,b'b}& (u_s)_{a'a}\\
	    (u_s^\dagger)_{b'b}  & (u_s) 
\end{array}\right) 
\end{equation}
for $s\ne 0$, 
while for $s=0$ the subamplitude $t_s$ has the standard three 
components as  mentioned above
\begin{equation}
{\bf t}_s=\matrix{(t_s)_{a'a,a'b}}\equiv (u_{a'})_{a,b} \, .
\end{equation}

The subamplitudes are solutions of the equation 
for the subsystem dynamics
\begin{mathletters}
\begin{eqnarray}
(t_{s})_{a'a,b'b}&=&(v_{s})_{a'a,b'b}+
         \sum_{c'(\subset s)}\sum_{c(\subset c')}
           (v_{s})_{a'a,c'c}G_0t_cG_0(t_{s})_{c'c,b'b}
          +(v_{s})_{a'a,-}g_0(t_{s})_{-,b'b} \, ,\\
(t_{s})_{-,b'b}&=&(v_{s})_{-,b'b}+
         \sum_{c'(\subset s)}\sum_{c(\subset c')}
           (v_{s})_{-,c'c}G_0t_cG_0(t_{s})_{c'c,b'b}
          +(v_{s})_{-,-}g_0(t_{s})_{-,b'b} \, ,\\
(t_{s})_{a'a,-}&=&(v_{s})_{a'a,-}+
         \sum_{c'(\subset s)}\sum_{c(\subset c')}
           (v_{s})_{a'a,c'c}G_0t_cG_0(t_{s})_{c'c,-}
          +(v_{s})_{a'a,-}g_0(t_{s})_{-,-} \, ,\\
(t_{s})_{-,-}&=&(v_{s})_{-,-}+
         \sum_{c'(\subset s)}\sum_{c(\subset c')}
           (v_{s})_{-,c'c}G_0t_cG_0(t_{s})_{c'c,-}
          +(v_{s})_{-,-}g_0(t_{s})_{-,-} \, ,
\end{eqnarray}
\label{subdynamics}
\end{mathletters}

which can be explicitly written as
\begin{mathletters}
\begin{eqnarray}
 (u_s)_{a'a,b'b}&=&G_0^{-1}\bar\delta_{ab}\delta_{a'b'}
		 + \sum_{c'(\subset s)}\sum_{c(\subset c')}
                  \bar\delta_{ac}\delta_{a'c'} t_c G_0 (u_s)_{c'c,b'b}
		 +(f_s)_{a'a} g_0 (u_s^\dagger)_{b'b}\, ,
    \\
    (u_s^\dagger)_{a'a}&=&(f_s^\dagger)_{a'a}+{\cal V}_sg_0
			 (u_s^\dagger)_{a'a}+
                   \sum_{c'(\subset s)}\sum_{c(\subset c')}
                         (f_s^\dagger)_{c'c}
			 G_0t_cG_0(u_s)_{c'c,a'a}\, ,
 \\
    (u_s)_{a'a}&=&(f_s)_{a'a}+
         \sum_{c'(\subset s)}\sum_{c(\subset c')}
    \bar\delta_{ac}\delta_{a'c'}t_cG_0(u_s)_{c'c}+
    (f_s)_{a'a}g_0(u_s)\, ,
    \\
    (u_s)&=&{\cal V}_s+{\cal V}_s g_0(u_s)+
         \sum_{c'(\subset s)}\sum_{c(\subset c')}
    {(f_s^\dagger)}_{c'c}G_0t_cG_0(u_s)_{c'c}\, .
   \end{eqnarray}
\label{ABsub}
\end{mathletters}

One can directly compare the structure of these equations
with the previously discussed AB equations, Eqs.~(\ref{AB0}). 
They are obviously similar, 
the former being the dynamical equation for the whole system,
the latter carrying the information for the internal dynamics
with respect to the partition $s$.
In Eqs.~(\ref{ABsub}) a careful disentanglement has been
made of which components contribute within the same subsystem,
according to the scheme illustrated in Tab.~\ref{tab3}.
      We observe that for each partition $s\ne 0$ the no-pion
sector acts as a doorway state and couples together two different
two-cluster partitions $a'$ of the four-body sector.
The operators $(f_s)_{a'a}$ and $(f^\dagger_s)_{b'b}$ are fundamental in 
this sense, since without these the two-cluster partitions of the
four-body sector would remain uncoupled (as happens in the 
standard four-body theory).

When $s=0$ the subamplitude is a genuine
four-body subamplitude, identified by  
one single two-cluster partition of the four-body system.
Nevertheless, we prefer to write the equation for the $s=0$ subamplitude 
as follows
\begin{eqnarray}
 (u_s)_{a'a,b'b}&=&G_0^{-1}\bar\delta_{ab}\delta_{a'b'}
		 +\sum_{c'(\subset s)}\sum_{c(\subset c')}
                  \bar\delta_{ac}\delta_{a'c'} t_cG_0
		  (u_s)_{c'c,b'b}.
\label{s=0subampl}
\end{eqnarray}

\leftline

]

Clearly, this is not the most simple way to write a standard AGS equation
(in presence of a spectator particle),
however it does correspond to the standard AGS equation 
since only the [$(NNN)\ \pi$] 
partition is relevant for $s=0$ 
(hence $a'$ = $b'$ = $c'$ = [$(NNN)\ \pi$]). 
The form given by Eq.~(\ref{s=0subampl})
has the advantage that treats the $s=0$ subamplitude
with the same formalism introduced to describe
the much more complex  $s\ne 0$ subamplitudes.

%\leftline
%\setlength{\textwidth}{7in}
%\setlength{\textheight}{9.5in}

%@twocolumntrue\endcsname

\begin{figure}[t]
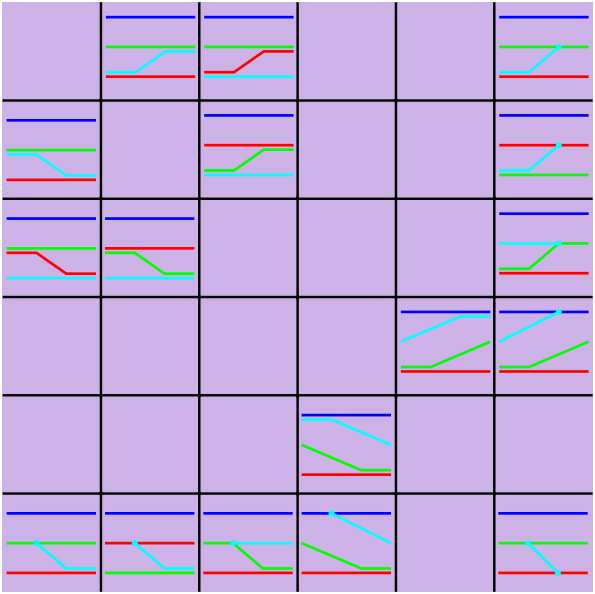

\begin{center}
\Spicture{3.50in}{dfpd98th22diag1.ps}{}
\end{center}
\caption[*]{Disconnected three-cluster exchange diagrams ${\bf z}_s$, for $s=1$.
These diagrams contribute to the interaction between nucleons ``2" and ``3",
(green and red lines, respectively).
The blu line (nucleon ``1") is always disconnected from the green 
and red ones, for any iteration of the diagrams belonging to this set.
The pale blue line represents the pion.\label{fig1}}
\end{figure}

%]

Up to now we have discussed the partition modes of the $\pi$NNN-NNN
system into two clusters and have given the corresponding subsystem equations.
We show now that the channel interaction $v_s$ satisfies a sum-rule 
property. For convenience, we discuss separately the 
effect of the sum-rule in the various sectors.

In the four-body sector, the driving term (total interaction)
of the AB equations is a matrix potential with components 
ranging within the 6 three-cluster partitions of the system
$V^{(3)}_{ab}=G_0^{-1}\bar\delta_{ab}$.
In the same sector the channel interaction has a structure
which is conceptually more complicated, since for each partition
$s$ the potential in the four-body sector
is a matrix potential ranging between all the possible chains of partitions
corresponding to each~$s$ :
$(v_s)_{a'a,b'b}=G_0^{-1}\bar\delta_{ab}\delta_{a'b'}\delta_{a,b\subset a'}
\delta_{a'\subset s}$.
In particular for each partition with
$s\ne 0$ we have five chains while for $s=0$ there are three chains.
The total corresponds to the 18 Yakubovsk\u{\i} components necessary 
for the complete dynamical description of four-body states.
We observe that the following sum rule holds
\begin{equation}
(V^{(3)})_{ab}=\sum_{s=0}^3\sum_{a',b'(\subset s)}(v_s)_{a'a,b'b}.
\end{equation}
This can be easily demonstrated once it has been realized  
that the right-hand term can be rewritten
as $\sum_{a'}G_0^{-1}\bar\delta_{ab}\delta_{a,b\subset a'}$.

Similarly, for the interaction operators
connecting the four-body and the three-nucleon sectors,
we observe the following sum rules
\begin{mathletters}
\begin{eqnarray}
F_a=\sum_{s}\sum_{a'(\subset s)}(f_s)_{a'a}, \\
F^\dagger_a=\sum_{s}\sum_{a'(\subset s)}(f_s^\dagger)_{a'a}.
\end{eqnarray}
\end{mathletters}
They both come from the identity
\begin{eqnarray}
%\sum_{i=1}^3
\bar\delta_{ia} =
\sum_s\sum_{a'}
%\sum_{i=1}^3
\bar\delta_{ia}\delta_{i,a\subset a'}
\delta_{a'\subset s},
\end{eqnarray}
which can be demonstrated by observing (from Tab.~\ref{tab3})
that a partition $a'$ corresponds to one single subsystem 
$s$ and a pair of {\em different}
three-cluster partitions ${i,a}$ corresponds to one single 
two-cluster partition $a'$. Furthermore, we observe that
the $s=0$ contribution to the sum over $s$ is identically null 
since there are no pion-nucleon pairs 
%(denoted in the formalism 
%with the index $i$ while $a$ denotes a generic pair)
which can be identified from the sequential break-up
of the $\pi\ (NNN)$ partition.

Finally, in the no-pion sector, ${\cal V}$ represents
the sum over all the pair interactions amongst the 
three nucleons, 
\begin{equation}
{\cal V}=\sum_{s} {\cal V}_s ,
\end{equation}
having assumed that only two-body NN potentials are given as 
input. 
The sum over the three $s$ components (from 1 to 3)
saturates the total interaction in the three-nucleon
sector (obviously the $s$=0 case does not contribute here as well as 
in the vertex interactions).

We summarize the results obtained so far:

(I) Our starting point is given by the AB equations 
which have been symbolically rewritten as a matrix 
LS equation
connecting all the three-cluster partitions (in both sectors) of 
the system.
\begin{equation}
{\bf T}^{(3)}={\bf V}^{(3)}+{\bf V}^{(3)}{\bf G}_0^{(3)}{\bf T}^{(3)}.
\label{(I)}
\end{equation}
 
(II) We have introduced the dynamical equations
for the subamplitudes. Since we have already expressed these equations
in detail [in Eqs.~(\ref{ABsub})], we rewrite the same equations
in a more compact matrix form, namely
\begin{equation}
{\bf t}_s={\bf v}_s+{\bf v}_s{\bf G}_0^{(3)}{\bf t}_s.
\label{(II)}
\end{equation}
%where we have put in evidence only the role played 
%by the two-cluster partition indices of the four-body sector, 
%${a',b',c'}$, which are given once the two cluster partition
%of the whole system $s$ has been fixed. Written in the present form,
%the equation has to be taken with great care. 
It has to 
be recalled that only when $s\ne$0 there is a direct coupling to 
the three-nucleon sector.
%In this last case, the equation overflows into the three--nucleon sector,
%where $s$ directly represents the partition of the system, and obviously 
%the chain-of-partition indices of the four-body systems
%are not relevant any more.
%representing clusters in the four-body sector
%no longer exist. 
%Finally, when $s\ne 0$, also in the 
%intermediate product $\sum_{c'}(v^{s})_{a'c'}G_0^{(3)}(t_{s})_{c'b'}$,
%one must include the contribution from the no-pion sector,
%as expressed clearly in Eq.~(\ref{ABsub}).
The operators involved in Eq.~(\ref{(II)}) act in a conceptually more
complex space, if compared to the three-cluster partition space of
${\bf T}^{(3)}$, ${\bf V}^{(3)}$, and ${\bf G}_0^{(3)}$, 
since it involves the chain-of-partition labelling
of the Yakubovsk\u{\i} approach. Therefore, care must be taken 
in considering the operatorial 
\twocolumn[\hsize\textwidth\columnwidth\hsize\csname
@twocolumnfalse\endcsname
product
${\bf v}_s{\bf G}_0^{(3)}{\bf t}_s$ since the operators
are defined in different spaces, as can be directly seen by
inspection of the detailed formulas previously reported.

(III) Within this formalism, we can collect the three sum rules 
previously discussed in a more general and compact sum rule
%\leftline
%

\begin{eqnarray}
\label{(III)}
{\bf V}^{(3)}& \equiv& \left(
\begin{array}{ll}          (V^{(3)})_{ab}&
                           (V^{(3)})_{a,-}\\
                           (V^{(3)})_{-,b}&
                           (V^{(3)})_{-,-}
        \end{array}\right)
%\\ \nonumber
= \left(
\begin{array}{cc} \sum_{s}\sum_{a',b'(\subset s)}(v_{s})_{a'a,b'b}&
                  \sum_{s}\sum_{a'(\subset s)}(v_{s})_{a'a,-}\\
                  \sum_{s}\sum_{b'(\subset s)}(v_{s})_{-,b'b}&
                  \sum_{s}(v_{s})_{-,-}
        \end{array}\right).
\end{eqnarray}

%where the components referring to the four-body sector, 
%in both sides of the equation, have to be considered as 
%matrix operators spanning the three-cluster partition
%of the system.
%]

(IV) We now can proceed in analogy with the methods
developed in standard N-body theory, namely we introduce the new 
unknowns U, with the following definition:

%
%\leftline
%@twocolumnfalse\endcsname
%
\begin{mathletters}
\begin{eqnarray}
\nonumber
(T^{(3)})_{ab}&=&\sum_{s} \sum_{a',b'(\subset s)}(t_{s})_{a'a,b'b}
%\\ \nonumber & &        
+\sum_{s,s'} \sum_{a',c'(\subset s)}\sum_{d',b'(\subset s')}
  \sum_{c(\subset c')} \sum_{d(\subset d')}
         (t_{s})_{a'a,c'c}G_0 t_c G_0 (U_{s,s'})_{c'c,d'd}
         G_0 t_d G_0 (t_{s'})_{d'd,b'b} \\
\nonumber
& &+\sum_{s,s'} \sum_{a',c'(\subset s)}\sum_{b'(\subset s')}
  \sum_{c(\subset c')}
         (t_{s})_{a'a,c'c}G_0 t_c G_0 (U_{s,s'})_{c'c,-}
         g_0 (t_{s'})_{-,b'b} \\
\nonumber
& &+\sum_{s,s'} \sum_{a'(\subset s)}\sum_{d',b'(\subset s')}
   \sum_{d(\subset d')}
         (t_{s})_{a'a,-} g_0 (U_{s,s'})_{-,d'd}
         G_0 t_d G_0 (t_{s'})_{d'd,b'b} \\
& &+\sum_{s,s'} \sum_{a'(\subset s)}\sum_{b'(\subset s')}
         (t_{s})_{a'a,-}g_0 (U_{s,s'})_{-,-}
         g_0 (t_{s'})_{-,b'b} \\
\nonumber
(T^{(3)})_{a-}&=&\sum_{s} \sum_{a'(\subset s)}(t_{s})_{a'a,-}
%\\ \nonumber & &        
+\sum_{s,s'} \sum_{a',c'(\subset s)}\sum_{d'(\subset s')}
  \sum_{c(\subset c')}\sum_{d(\subset d')}
         (t_{s})_{a'a,c'c}G_0 t_c G_0 (U_{s,s'})_{c'c,d'd}
         G_0 t_d G_0 (t_{s'})_{d'd,-} \\
\nonumber
& &+\sum_{s,s'} \sum_{a',c'(\subset s)}
  \sum_{c(\subset c')}
         (t_{s})_{a'a,c'c}G_0 t_c G_0 (U_{s,s'})_{c'c,-}
         g_0 (t_{s'})_{-,-} \\
\nonumber
& &+\sum_{s,s'} \sum_{a'(\subset s)}\sum_{d'(\subset s')}
  \sum_{d(\subset d')}
         (t_{s})_{a'a,-} g_0 (U_{s,s'})_{-,d'd}
         G_0 t_d G_0 (t_{s'})_{d'd,-} \\
& &+\sum_{s,s'} \sum_{a'(\subset s)}
         (t_{s})_{a'a,-}g_0 (U_{s,s'})_{-,-}
         g_0 (t_{s'})_{-,-} \\
\nonumber
(T^{(3)})_{-b}&=&\sum_{s} \sum_{b'(\subset s)}(t_{s})_{-,b'b}
%\\ \nonumber & &        
+\sum_{s,s'} \sum_{c'(\subset s)}\sum_{d',b'(\subset s')}
  \sum_{c(\subset c')}\sum_{d(\subset d')}
         (t_{s})_{-,c'c}G_0 t_c G_0 (U_{s,s'})_{c'c,d'd}
         G_0 t_d G_0 (t_{s'})_{d'd,b'b} \\
\nonumber
& &+\sum_{s,s'} \sum_{c'(\subset s)}\sum_{b'(\subset s')}
  \sum_{c(\subset c')}
         (t_{s})_{-,c'c}G_0 t_c G_0 (U_{s,s'})_{c'c,-}
         g_0 (t_{s'})_{-,b'b} \\
\nonumber
&  &+\sum_{s,s'} \sum_{d',b'(\subset s')}
  \sum_{d(\subset d')}
         (t_{s})_{-,-} g_0 (U_{s,s'})_{-,d'd}
         G_0 t_d G_0 (t_{s'})_{d'd,b'b} \\
& &+\sum_{s,s'} \sum_{b'(\subset s')}
         (t_{s})_{-,-}g_0 (U_{s,s'})_{-,-}
         g_0 (t_{s'})_{-,b'b} \\
\nonumber
(T^{(3)})_{-,-}&=&\sum_{s} (t_{s})_{-,-}
%\\ \nonumber & &        
+\sum_{s,s'} \sum_{c'(\subset s)}\sum_{b'(\subset s')}
  \sum_{c(\subset c')}\sum_{d(\subset d')}
         (t_{s})_{-,c'c}G_0 t_c G_0 (U_{s,s'})_{c'c,d'd}
         G_0 t_d G_0 (t_{s'})_{d'd,-} \\
\nonumber
& &+\sum_{s,s'} \sum_{c'(\subset s)}
  \sum_{c(\subset c')}
         (t_{s})_{-,c'c}G_0 t_c G_0 (U_{s,s'})_{c'c,-}
         g_0 (t_{s'})_{-,-} \\
\nonumber
& &+\sum_{s,s'} \sum_{a'(\subset s)}\sum_{d'(\subset s')}
  \sum_{d(\subset d')}
         (t_{s})_{a'a,-} g_0 (U_{s,s'})_{-,d'd}
         G_0 t_d G_0 (t_{s'})_{d'd,-} \\
& &+\sum_{s,s'} 
         (t_{s})_{-,-}g_0 (U_{s,s'})_{-,-}
         g_0 (t_{s'})_{-,-} .
\end{eqnarray}
\label{(IV)}
\end{mathletters}
]
%As usual each of the operators defined in the above expression
%is acting the all the partitions of the system into three clusters,
%which means that in the four-body sector the operators
%are endowed of an additional index referring to such partitions.

%\leftline

\twocolumn[\hsize\textwidth\columnwidth\hsize\csname
@twocolumnfalse\endcsname
Now, we substitute Eqs.~(\ref{(III)}, \ref{(IV)}) into Eq.~(\ref{(I)}), 
 and use repeatedly
Eq.~(\ref{(II)}). We find that 
\begin{mathletters}
\begin{eqnarray}
\nonumber
(U_{s,s'})_{a'a,b'b} &=& 
(G_0t_aG_0)^{-1}\delta_{ab}
\left(\bar\delta_{ss'}+\delta_{ss'}\bar\delta_{a'b'}\right)\\
\nonumber
&& +
\sum_{s''}\sum_{c',d'(\subset s'')} \sum_{d(\subset d')}
\left(\bar\delta_{ss''}+\delta_{ss''}\bar\delta_{a'c'}\right)
(t_{s''})_{c'a,d'd }G_0 t_d G_0 
(U_{s'',s'})_{d'd,b'b}\\
&& + \sum_{s''}\sum_{c'(\subset s'')} 
\left(\bar\delta_{ss''}+\delta_{ss''}\bar\delta_{a'c'}\right)
(t_{s''})_{c'a,- } g_0 
(U_{s'',s'})_{-,b'b} 
\label{4Bscatt}\\
\nonumber
(U_{s,s'})_{-,b'b} &=&
\sum_{s''}\sum_{d'(\subset s'')} \sum_{d(\subset d')}
\left(\bar\delta_{ss''}\right)
(t_{s''})_{-,d'd }G_0 t_d G_0 
(U_{s'',s'})_{d'd,b'b}\\
&& + \sum_{s''} 
\left(\bar\delta_{ss''}\right)
(t_{s''})_{-,- } g_0 
(U_{s'',s'})_{-,b'b} \\
\nonumber
(U_{s,s'})_{a'a,-} &=&
\sum_{s''}\sum_{c',d'(\subset s'')} \sum_{d(\subset d')}
\left(\bar\delta_{ss''}+\delta_{ss''}\bar\delta_{a'c'}\right)
(t_{s''})_{c'a,d'd }G_0 t_d G_0 
(U_{s'',s'})_{d'd,-}\\
&& + \sum_{s''}\sum_{c'(\subset s'')} 
\left(\bar\delta_{ss''}+\delta_{ss''}\bar\delta_{a'c'}\right)
(t_{s''})_{c'a,- } g_0 
(U_{s'',s'})_{-,-} \\
\nonumber
(U_{s,s'})_{-,-} &=& 
(g_0)^{-1}
\left(\bar\delta_{ss'}\right) + \sum_{s''} 
\left(\bar\delta_{ss''}\right)
(t_{s''})_{-,- } g_0 
(U_{s'',s'})_{-,-} 
\\
&& +
\sum_{s''}\sum_{d'(\subset s'')} \sum_{d(\subset d')}
\left(\bar\delta_{ss''}\right)
(t_{s''})_{-,d'd }G_0 t_d G_0 
(U_{s'',s'})_{d'd,-} \, .
\label{3Nscatt}
\end{eqnarray}
\label{finalresult}
\end{mathletters}
\leftline
]
% commenti vari all'equazione
It must be observed that there is always a relation
between the chains of partitions of the four-body sector, $\{a'a\}$,
and the two-cluster partitions $s$, since for a given 
$s$, the allowed partitions ($a'\subset s$) are enlisted in Tab.~\ref{tab3}.
Keeping this in mind, it is obvious that
$\left(\bar\delta_{ss'}+\delta_{ss'}\bar\delta_{a'b'}\right)
=\bar\delta_{a'b'}$.

These four coupled equations represent the main theoretical result
of the paper. The first two equations couple together
four-body scattering and pion absorption, 
while the last two couple three-nucleon scattering with 
pion production. 
The equations decouple into ordinary four- and three-body
equations if we switch off the couplings between the three 
and four-particle channels, however this is much less obvious 
than the corresponding decoupling for the simpler $\pi$NN system.
To show how this happens one must first observe that the four two-cluster partitions
of the whole system decouple into the seven two-cluster partition
of the four-body sector, plus the three two-cluster partitions
of the three-nucleon sector. Moreover all the production/absorption 
amplitudes vanish, and therefore Eq.~(\ref{3Nscatt})
changes into the standard three-component 
AGS equation, and Eq.~(\ref{4Bscatt}) becomes 
precisely the standard 18-component GS equation.
With the pion-nucleon vertex interaction switched on, we have instead
a new 21-component equation which is remarkably different in structure.

In the following we intend to discuss the properties of this set of equations
in the light of the quasiparticle interpretation. Then we will derive the 
corresponding bound-state equation and finally give the rules for calculating
the collision amplitudes for rearrangement and break-up processes.

\section{The quasiparticle formalism}
\label{quasiparticle}

The introduction of the
quasiparticle formalism is in principle not indispensable, since 
direct solutions of multivariable few-body-type integral equations
are possible
by resorting to the 
nowadays available computational tools. 
The historical reason for introducing the quasiparticle
method is that it reduces by one unit 
the dimensionality of the multiparticle equation
whenever the method is applied. By repeated applications
of the method, one reduces the problem to the solution
of a two-cluster multiparticle equation in one single variable,
after angular momentum decomposition. 
However, the quasiparticle or separable method does not represent only a 
converging approximation scheme
but it allows also to reinterpret the previously obtained equations
in a physically more transparent way, and by translating 
the theory in terms of coalescence diagrams, it allows
to exhibit diagrammatically the connected-kernel 
properties of the final equations.
%Separable methods, when are applied with minimum care, 
%describe accurately the polar behaviour for the subamplitudes, 
%and provide directly the reaction amplitudes for cluster processes.

To introduce the quasiparticle formalism, we derive first
the amplitude for the fully unclusterized reaction process.
This corresponds to the 4 to 4 amplitude, denoted by $T(1|1)$, 
describing the process 
%(very unlikely, from the experimental point of view) 
of a free collision of the four particles. 
The amplitude for this process is linked to the 
Afnan-Blankleider amplitudes previously defined, ${\bf T}^{(3)}$
(we remind that the AB amplitudes for the $\pi$NNN-NNN
system refer to all the 3 to 3 processes).
To obtain this link, we resort to Ref.\cite{ccs}
where the AB theory for the $\pi$NNN-NNN system
has been discussed within the diagrammatic approach.

As shown in Ref.\cite{ccs},
if we apply the Last-Cut Lemma 

\twocolumn[\hsize\textwidth\columnwidth\hsize\csname
@twocolumnfalse\endcsname
to the $4\leftarrow 4$ amplitude
we obtain
\begin{equation}
T(1|1)= T(1|0)g_0T(0|1)_1 + T(1|1)_1\, ,
\end{equation}
while applying the First-Cut Lemma to the $4\leftarrow 3$
amplitude yields
\begin{equation}
T(1|0)= T(1|0)_1\left(1+g_0T(0|0)\right) \, .
\end{equation}
The subscript ``1" denotes that the given amplitude 
contains at least one pion in all the intermediate states.

Similar assumptions for $T(0|1)_1$ and $T(1|0)_1$ yield
(to the lowest order, see Eqs. (2.6), (2.8), and (2.10) of 
Ref.\cite{ccs})
\begin{eqnarray}
T(1|0)_1 = (\sum_i f^{(o)}_i) + T(1|1)_1 G_0 (\sum_i f^{(o)}_i)  \\
T(0|1)_1 = (\sum_i{f^{(o)}_i}^\dagger) + 
(\sum_i{f^{(o)}_i}^\dagger) G_0 T(1|1)_1 \, ,
\end{eqnarray}
and from these last equations we obtain
\begin{eqnarray}
T(1|1)= T(1|1)_1 + (1+ T(1|1)_1G_0)
(\sum_i f^{(o)}_i)
(g_0+g_0T(0|0)g_0) 
(\sum_i{f^{(o)}_i}^\dagger)
(1+ G_0T(1|1)_1).
\end{eqnarray}
If we identify $T(1|1)_1$ with the standard 4-body 4 to 4 amplitude,
$T(1|1)_1=U_{00}$,
we can use the relations connecting the various 
AGS amplitudes, 
\begin{eqnarray}
U_{00}=U_{0i}(1+G_0t_i)-G_0^{-1}\\
U_{00}=(1+t_iG_0)U_{i0}-G_0^{-1}
\end{eqnarray}
By substituting the two expressions in the previous formula
we get 
\begin{equation}
T(1|1)= %\sum_a t_a +\sum_{a,b} t_a G_0 U_{ab} G_0 t_b %opp 
U_{00}
+
\sum_{ij} U_{0i} G_0 f_i (g_0 + g_0 T(0|0) g_0) f^\dagger_j U_{j0} 
\end{equation}
and recalling that
\begin{eqnarray}
U_{00} &=& \sum_a t_a +\sum_{a,b} t_a G_0 U_{ab} G_0 t_b \\
U_{0i} &=& G_0^{-1}+\sum_{c=1,6} t_cG_0U_{ci}\\
U_{i0} &=& G_0^{-1}+\sum_{c=1,6} U_{ic}G_0t_c
\end{eqnarray}
we obtain 
\begin{eqnarray}
\nonumber
T(1|1) &=&
\sum_a t_a +\sum_{a,b} t_a G_0 U_{ab} G_0 t_b 
+ \sum_{aijb} t_aG_0U_{ai} 
f_i (g_0 + g_0 T(0|0) g_0) f^\dagger_j 
U_{jb}G_0t_b +\sum_{aij} t_aG_0U_{ai} 
f_i (g_0 + g_0 T(0|0) g_0) f^\dagger_j 
\\ &&
+
\sum_{ijb}  
f_i (g_0 + g_0 T(0|0) g_0) f^\dagger_j 
U_{jb}G_0t_b
% \\ &&
+
\sum_{ij}  
f_i (g_0 + g_0 T(0|0) g_0) f^\dagger_j  \, .
\end{eqnarray}
By the use of the AGS equations (see Ref.\cite{ccs}, 
pag. 1238-1240), it is possible to directly express the above 
amplitude in terms of the AB amplitudes for the three--cluster
partitions of the system, thereby obtaining the final result
\begin{eqnarray}
T(1|1) &=&
\sum_a t_a + 
\sum_{ij}  
f_i g_0 f^\dagger_j + \sum_{a,b} t_a G_0 (T^{(3)})_{ab} G_0 t_b 
\nonumber\\
&&+
\sum_{aj} t_aG_0 (T^{(3)})_{a-} g_0 f^\dagger_j 
+
\sum_{ib}  
f_i g_0 (T^{(3)})_{-b} G_0t_b +
\sum_{ij}  
f_i g_0 (T^{(3)})_{-,-} 
 g_0 f^\dagger_j .
\label{t44}
\end{eqnarray}

\vskip -5 truecm
]
$\ $
\eject\newpage 

$\ $
\eject\newpage 

\twocolumn[\hsize\textwidth\columnwidth\hsize\csname
@twocolumnfalse\endcsname
It must be observed that in previous studies 
\cite{cc94ab,cc97} the second, and the three last
terms were missing in the reported expressions 
for the fully unclusterized amplitude $T{(1|1)}$. 
In particular, the simplest pole-type diagrams
$\sum_{ij}  
f_i g_0 f^\dagger_j $ were not considered in that approach.
We introduce now the quasiparticle method.
%\rightline
%\newpage
According to this method, the two--body t-matrix are represented
by means of the separable expansion,
\begin{equation}
t_a(z)=|a^{(3)}(z)>\tau^{(3)}_a(z)<a^{(3)}(z)|.
\label{rank-one}
\end{equation}
When calculating the matrix-element of this operator
in the four-body space, we obtain
%\nopagebreak
%]
%\nopagebreak
\begin{equation}
\langle {\bf p} {\bf q_1} {\bf q_2}
| t_a 
|{\bf p'} {\bf q'_1} {\bf q'_2}\rangle
=\delta({\bf q'_1-q_1}) \delta({\bf q'_2-q_2})
\langle {\bf p}|a^{(3)}(z-\Delta)\rangle \tau_a^{(3)}({z-\Delta})
\langle a^{(3)}(z-\Delta)| {\bf p' } \rangle \, ,
\end{equation}

\leftline 
]
where it is assumed that ${\bf p}$, is the relative momentum
of the pair $a$, while ${\bf q_1}$,  ${\bf q_2}$ are
the Jacobi coordinates for the
two spectators and the c.m. of the pair 
(considered in toto as a three-body system),
and $z-\Delta({\bf q_1},{\bf q_2})$ the kinetic energy of the pair 
$a$ with respect to its c.m. 

Here, for simplicity, we have assumed a rank-one structure, 
but the extension of the formalism to higher ranks is 
straightforward, although practical extensions 
might require a major computational work.
Depending on the specific separable expansion method,
the states may or may not depend on the parametric energy, $z$.
Moreover, $<a^{(3)}(z)|$ does not necessarily have to be 
the adjoint of $|a^{(3)}(z)>$; for instance, in case of Weinberg states
a possible choice is
$<a^{(3)}(z)| = {|a^{(3)}(z^*)>}^\dagger $, but depending on the normalization 
conventions other choices are also possible
\cite{racan}.

We have no reasons here for 
analyzing in detail 
the technical differences which characterize the variety of 
separable-expansion methods available in the 
Literature (for this we refer to Ref.~\cite{adikowa});
as long as they correctly reproduce the 
polar structure of the subsystem t-matrices
we generically denote all these methods as
``quasiparticle" approaches, although the quasiparticle idea
historically refers to the application in terms of Weinberg 
states~\cite{WEI}.

We note that the separable assumption affects only the 4-body space, given 
that the two-body t-matrices $t_a$ act within this space, and, by
means of the form Eq.~(\ref{rank-one}), the fully unclusterized amplitude
becomes (omitting the superscript ``(3)" in the states $|a\rangle$)
\begin{eqnarray}
T(1|1) &=&
\sum_a |a>\tau_a<a| 
+ \sum_{ij}  f_i g_0 f^\dagger_j 
\nonumber\\&&
+ \sum_{ab} |a>\tau_a X^{(3)}_{ab} \tau_b<b| 
+ \sum_{aj} |a>\tau_a X^{(3)}_{a} g_0 f^\dagger_j 
\nonumber\\&&
+ \sum_{ib}  f_i g_0 {X^\dagger_{b}}^{(3)} \tau_b<b| 
+\sum_{ij}  f_i g_0 X^{(3)} g_0 f^\dagger_j 
\nonumber\\&&
\end{eqnarray}
where the folded amplitudes
are given 
according to %the equations
%\rightline
\newpage
\begin{mathletters}
\begin{eqnarray}
X_{ab}^{(3)}&=&<a|G_0 (T^{(3)})_{ab} G_0 |b>\\
X_{a}^{(3)}&=&<a|G_0 (T^{(3)})_{a-} \\
{X^\dagger_{b}}^{(3)}&=&(T^{(3)})_{-b} G_0 |b>\\
X^{(3)}&\equiv& (T^{(3)})_{-,-} \, .
\end{eqnarray}
\end{mathletters}
In the ${\bf X}^{(3)}$ amplitudes 
the variable describing the internal structure
of the pair has been integrated over,
thereby reducing the dimensionality of the corresponding 
dynamical equation. Such quasiparticle equation
for the ${\bf X}^{(3)}$ amplitudes has been given in
Eq.~(2.6) of Ref.\cite{AM}. However it is known
that the equation is not connected for the pion-three-nucleon problem
\cite{AM,ccs}.
%and this can be understood with a very simple argument:
%The equation  for the $X^{(3)}$ amplitudes contain 
%as a subset all the conventional four--body diagrams,
%and summing the complete perturbation series for this 
%restricted set of diagrams, as is well known, leads to 
%non-connected-kernel four-body equations.  
%\rightline
%\newpage
\begin{figure}[t]
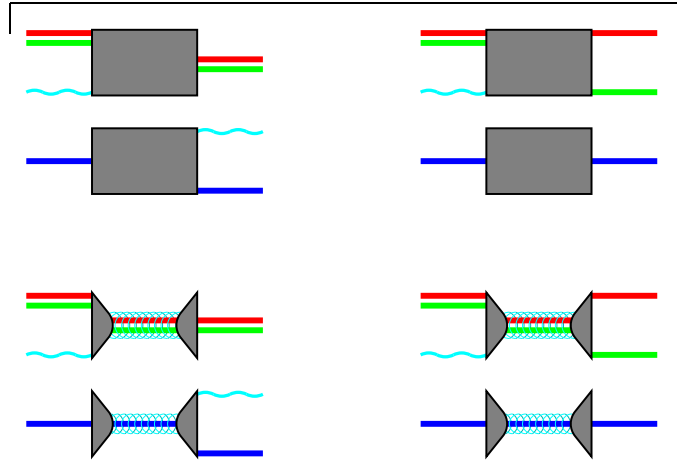

\vskip 0.3 truecm
\begin{center}
\Spicture{4.3in}{dfpd98th22diag2.ps}{}
\end{center}
\vskip -6.0 truecm
\caption[*]{Examples of disconnected three-cluster amplitudes, for $s=1$.
The two diagrams on top of the figure represent the subamplitude 
${\bf x}_s$,
with $s=1$. In particular, the case $(x_s)_{a'a,b'b}$
with $a'=b'$ and $a\neq b$ has been chosen for the top-left diagram, while 
the top-right diagram represents the production subamplitude 
$(x_s)_{a'a,-}$. The corresponding diagrams on the bottom side 
denote the very same amplitudes in the quasiparticle formalism. 
Here, the intermediate propagation of the multiparticle 
two-fragment partition is exhibited by drawing the nucleonic lines 
surrounded by a pionic concentric line. For $s=1$ the three possible 
intermediate two-cluster components are $[(\pi N_2 N_3) N_1]$, 
$[(N_2 N_3) (N_1 \pi)]$, and  $[(N_2 N_3) N_1]$. \label{fig2}}
\leftline
\end{figure}

\clearpage

\twocolumn[\hsize\textwidth\columnwidth\hsize\csname
@twocolumnfalse\endcsname

We solve the problem by introducing the representation given in 
Eq.~(\ref{(IV)}) which allows to express the three-cluster partition
amplitudes in terms of the new quantities $t_s$ and $U_{ss'}$

\begin{eqnarray}
\nonumber
T(1|1) &=&
\sum_a t_a + 
\sum_{ij}  
f_i g_0 f^\dagger_j + 
\sum_s\sum_{a',b'(\subset s)} \sum_{a(\subset a')} \sum_{b(\subset b')} 
t_a G_0 (t_s)_{a'a b'b} G_0 t_b
\\ \nonumber 
&&+
\sum_{ss'}
\sum_{a',c'(\subset s)}
\sum_{b',d'(\subset s')} 
\sum_{a(\subset a')}
\sum_{c(\subset c')} 
\sum_{b(\subset b')}
\sum_{d(\subset d')}  
t_a G_0 (t_s)_{a'a, c'c} G_0 t_c G_0 (U_{ss'})_{c'c,d'd} G_0 t_d G_0 
(t_{s'})_{d'd, b'b} G_0 t_b\\ \nonumber
&&+
\sum_{ss'}
\sum_{a'(\subset s)}
\sum_{b',d'(\subset s')} 
\sum_{a(\subset a')} 
\sum_{b(\subset b')}
\sum_{d(\subset d')}  
t_a G_0 (t_s)_{a'a,-} g_0 (U_{ss'})_{-,d'd} G_0 t_d G_0 
(t_{s'})_{d'd, b'b} G_0 t_b\\ \nonumber
&&+
\sum_{ss'}
\sum_{a',c'(\subset s)}
\sum_{b'(\subset s')} 
\sum_{a(\subset a')}
\sum_{c(\subset c')} 
\sum_{b(\subset b')}  
t_a G_0 (t_s)_{a'a, c'c} G_0 t_c G_0 (U_{ss'})_{c'c,-} g_0  
(t_{s'})_{-,b'b} G_0 t_b\\ \nonumber
&&+
\sum_{ss'}
\sum_{a'(\subset s)}
\sum_{b'(\subset s')} 
\sum_{a(\subset a')} 
\sum_{b(\subset b')}  
t_a G_0 (t_s)_{a'a,-} g_0 (U_{ss'})_{-,-} g_0  
(t_{s'})_{-,b'b} G_0 t_b\\ \nonumber
&&+
\sum_s
\sum_{a'(\subset s)}
\sum_{a(\subset a')}
t_a G_0 (t_s)_{a'a,-} g_0 (\sum_j f_j^\dagger)\\ \nonumber
&&+
\sum_{ss'}
\sum_{a',c'(\subset s)}
\sum_{a(\subset a')}
\sum_{c(\subset c')}
\sum_{b'(\subset s')}
\sum_{b(\subset b')}
t_a G_0 (t_s)_{a'a,c'c} G_0 t_c G_0 
(U_{ss'})_{c'c,b'b} G_0 t_b G_0 (t_{s'})_{b'b,-} g_0 
 (\sum_j f_j^\dagger)\\ \nonumber
&&+
\sum_{ss'}
\sum_{a'(\subset s)}
\sum_{a(\subset a')}
\sum_{b'(\subset s')}
\sum_{b(\subset b')}
t_a G_0 (t_s)_{a'a,-} g_0 
(U_{ss'})_{-,b'b} G_0 t_b G_0 (t_{s'})_{b'b,-} g_0 
 (\sum_j f_j^\dagger)\\ \nonumber
&&+
\sum_{ss'}
\sum_{a',c'(\subset s)}
\sum_{a(\subset a')}
\sum_{c(\subset c')}
t_a G_0 (t_s)_{a'a,c'c} G_0 t_c G_0 
(U_{ss'})_{c'c,-} g_0 (t_{s'})_{-,-} g_0 
 (\sum_j f_j^\dagger)\\ \nonumber
&&+
\sum_{ss'}
\sum_{a'(\subset s)}
\sum_{a(\subset a')}
t_a G_0 (t_s)_{a'a,-} g_0 
(U_{ss'})_{-,-} g_0 (t_{s'})_{-,-} g_0 
(\sum_j f_j^\dagger)\\ \nonumber
&&+
\sum_s
\sum_{b'(\subset s)}
\sum_{b(\subset b')} 
(\sum_{i}  f_i) g_0 (t_s)_{-,b'b} G_0t_b \\ \nonumber
&&+
\sum_{ss'}
\sum_{b',d'(\subset s')}
\sum_{b(\subset b')}
\sum_{d(\subset d')} 
\sum_{c'(\subset s)}
\sum_{c(\subset c')} 
(\sum_{i}  f_i) g_0 (t_s)_{-,c'c} G_0t_cG_0
(U_{ss'})_{c'c,d'd} G_0 t_d G_0 (t_{s'})_{d'd,b'b} G_0t_b \\ \nonumber
&&+
\sum_{ss'}
\sum_{b',d'(\subset s')}
\sum_{b(\subset b')}
\sum_{d(\subset d')} 
(\sum_{i}  f_i) g_0 (t_s)_{-,-} g_0
(U_{ss'})_{-,d'd} G_0 t_d G_0 (t_{s'})_{d'd,b'b} G_0t_b \\ \nonumber
&&+
\sum_{ss'}
\sum_{b'(\subset s')}
\sum_{b(\subset b')} 
\sum_{c'(\subset s)}
\sum_{c(\subset c')} 
(\sum_{i}  f_i) g_0 (t_s)_{-,c'c} G_0t_cG_0
(U_{ss'})_{c'c,-} g_0 (t_{s'})_{-,b'b} G_0t_b \\ \nonumber
&&+
\sum_{ss'}
\sum_{b'(\subset s')}
\sum_{b(\subset b')} 
(\sum_{i}  f_i) g_0 (t_s)_{-,-} g_0
(U_{ss'})_{-,-} g_0 (t_{s'})_{-,b'b} G_0t_b \\ \nonumber
&&+
\sum_s
(\sum_i f_i)  g_0 (t_s)_{-,-} g_0  (\sum_j f^\dagger_j)\\ \nonumber
&&+
\sum_{ss'}
\sum_{c'(\subset s)}
\sum_{c(\subset c')} 
\sum_{d'(\subset s')}
\sum_{d(\subset d')}
(\sum_i f_i)  
g_0 (t_s)_{-,c'c} G_0t_cG_0 (U_{ss'})_{c'c,d'd} G_0 t_d G_0 (t_{s'})_{d'd,-} g_0
(\sum_j f^\dagger_j)\\ \nonumber
&&+
\sum_{ss'}
\sum_{d'(\subset s')}
\sum_{d (\subset d')}
(\sum_i f_i)  
g_0 (t_s)_{-,-} g_0 (U_{ss'})_{-,d'd} G_0 t_d G_0 (t_{s'})_{d'd,-} g_0
(\sum_j f^\dagger_j)\\ \nonumber
&&+
\sum_{ss'}
\sum_{c'(\subset s)}
\sum_{c(\subset c')} 
(\sum_i f_i)  
g_0 (t_s)_{-,c'c} G_0t_cG_0 (U_{ss'})_{c'c,-} g_0 (t_{s'})_{-,-} g_0
(\sum_j f^\dagger_j)\\
&&+
\sum_{ss'}
(\sum_i f_i)  
g_0 (t_s)_{-,-} g_0 (U_{ss'})_{-,-} g_0 (t_{s'})_{-,-} g_0
(\sum_j f^\dagger_j) \, .
\label{lunga}
\\ \nonumber
\end{eqnarray}
%\leftline
]
\newpage\clearpage

If we introduce at this point the quasiparticle expansion 
Eq.~(\ref{rank-one})
we obtain $T(1|1)$ expressed in terms of 
new folded amplitudes referring to the subsystem (or channel) dynamics
\begin{mathletters}
\begin{eqnarray}
(x_s)_{a'a,b'b}&=&<a|G_0 (t_s)_{a'a,b'b} G_0 |b>\\
(x_s)_{a'a,-}&=&<a|G_0 (t_s)_{a'a,-} \\
({x^\dagger_s})_{-,b'b}&=&(t_s)_{-,b'b} G_0 |b>\\
(x_s)_{-,-} &\equiv& (t_s)_{-,-} ,
\end{eqnarray}
\end{mathletters}
and to the total system
\begin{mathletters}
\begin{eqnarray}
({X}_{ss'})_{a'a,b'b}&=&<a|G_0 (U_{ss'})_{a'a,b'b} G_0 |b>\\
({X}_{ss'})_{a'a,-}&=&<a|G_0 (U_{ss'})_{a'a,-} \\
({X^\dagger_{ss'}})_{-,b'b}&=&(U_{ss'})_{-,b'b} G_0 |b>\\
(X_{ss'})_{-,-} &\equiv& (U_{ss'})_{-,-} .
\end{eqnarray}
\end{mathletters}
The corresponding expression of $T(1|1)$ in terms of ${\bf x}_s$
and ${\bf X}_{ss'}$ will be omitted for brevity but the derivation 
is quite obvious starting from
Eq.~(\ref{lunga}) : 
the quantities $U_{ss'}$, $t_s$, endowed
where appropriate with the Green's function $G_0$, are replaced 
by ${\bf X}_{ss'}$ and ${\bf x}_s$, respectively,
while the two-body t-matrix $t_a$ is substituted
with $\tau_a^{(3)}$. Finally
$\tau_a^{(3)}$ is further dressed with the state vector $|a^{(3)}>$ 
($<a^{(3)}|$) if the left (right) state refers to the asymptotic state
rather than to an intermediate state. 

The quasiparticle equation for the subsystem amplitudes  
can be immediately obtained by folding the equation
Eq.~(\ref{subdynamics}) between the states $<a|G_0$
and $G_0|b>$. The result is
\begin{equation}
{\bf x}_s={\bf z}_s+{\bf z}_s {\cal G}^{(3)} {\bf x}_s 
\label{x_s}
\end{equation}
where 
\begin{eqnarray}
{\bf z}_s
&&=
\left(
\begin{array}{ll}
(z_s)_{a'a,b'b}&(z_s)_{a'a,-}\\
({z}^\dagger_s)_{-,b'b}&(z_s)_{-,-}\\
\end{array}
\right)
\nonumber\\
&&=
\left(
\begin{array}{cc}
<a|G_0 |b>\delta_{a'b'}\bar\delta_{ab}
\delta_{ab\subset a'}
\delta_{a'\subset s}
&
<a|G_0 (f_s)_{a'a} \\
(f^\dagger_s)_{b'b} G_0 |b> &
{\cal V}_s
\end{array}
\right)\nonumber\\
\end{eqnarray}
and with the three-cluster (quasiparticle) propagator
given by
\begin{equation}
{\cal G}^{(3)}
= \left(\begin{array}{cc}
	\tau_a\delta_{ab} %\delta_{a'b'}questo lo tolgo previa def. del formalis.  
             & 0 \\
	0 & g_0
	\end{array}\right) .
\end{equation}
Within the same matrix formalism,
the solution of the equation for the subsystems
is represented as 
\begin{equation}
{\bf x}_s
=
\left(
\begin{array}{cc}
(x_s)_{a'a,b'b}&(x_s)_{a'a,-}\\
({x}^\dagger_s)_{-,b'b}&(x_s)_{-,-}\\
\end{array}
\right) ,
\end{equation}
where the elements (for each value of $s$)
are spanned by chain of partitions
in the one--pion sector, completed %orlati...%
with the additional component in the no-pion zone (in case $s\ne 0$),
in close analogy with the quantities ${\bf t}_s$.
This leads to 6$\times$6 matrices for $s\ne 0$,
while we have the standard 3$\times$3 matrix
for the $s=0$ partition.
Obviously, the same considerations previously
observed for ${\bf v}_s {\bf G}_0^{(3)} {\bf t}_s$
apply also for ${\bf z}_s {\cal G}^{(3)} {\bf x}_s$.

It might be useful to illustrate diagrammatically what 
the 6 components represent, {\it e.g.} for $s=1$, as has been 
done in Fig.~\ref{fig1}. Here the diagrams representing
the $z$-interaction, {\em i.e.} the driving term of 
Eq.~(\ref{x_s}), have been drawn.
The figure represents the diagrams in a square grid denoting
the 6$\times$6 interaction matrix. Both columns and rows are ordered
so that the first three elements represent the ($\pi$, $N_2$), ($\pi$, $N_3$),
and ($N_2$,$N_3$) pairs originating from the 
[($\pi$, $N_2$, $N_3$), $N_1$] two-cluster partition, the forth and fifth
elements represent the ($N_2$, $N_3$) and ($\pi$, $N_1$) pairs obtained
from the break-up of the second two-cluster partition, 
[($N_2$, $N_3$), ($\pi$, $N_1$)],
and finally the last element denotes the no-pion state 
with the three nucleons all disentangled.

In the bottom-right corner, one easily recognizes the two-nucleon 
OPE diagram, which is therefore extended in the present formulation
to embrace the entire set of diagrams shown by the figure.
As a matter of fact, for obvious reasons of simplicity
two diagrams have been omitted. One is a second OPE diagram,
similar to that already shown but with the opposite time ordering, 
and then (in the third row and last column) there should be
another diagram where the red and green lines (nucleons ``2" and ``3") 
are interchanged. It is clear that the same situation  occurs
in the symmetric case (third column and last row).

%%
%
%
%while in the subsequent Fig.~\ref{fig-x-mat},
%we draw the diagrams representing the $x$ amplitudes,
%for each matrix elements.  
%
%Osservazione: come rappresentare a mezzo diagrammi le sconnettivita' 
%delle equazioni per i sottosistemi?
%basta iterare il kernel ovvero i diagrammi z di figura {fig-z-mat}
%e far vedere che tutti i diagrammi iterati hanno lo stesso
%tipo di sconnettivita'
%o meglio, basta far vedere che ``z G x" sono ancora sconnessi.

In the same way as done for the subsystem dynamics, from 
Eq.~(\ref{finalresult})
it is possible to obtain the following equation
for the folded amplitudes referring to the entire system,  
which we write
as
\begin{equation}
{\bf X}_{ss'}={{\cal G}^{(3)}}^{-1}\bar\Delta_{ss'}
+\sum_{s''}\bar\Delta_{ss''}{\bf x}_{s''}{\cal G}^{(3)} {\bf X}_{s''s'}.
\label{step1}
\end{equation}

Here, we have introduced a new matrix-operator, $\bar\Delta$,
defined as follows
\begin{mathletters}
\begin{eqnarray}
({\bar\Delta}_{ss'})_{a'a,b'b}&\equiv&\delta_{ab}\bar\delta_{a'b'}
= \delta_{ab}\left(\bar\delta_{ss'}+
\delta_{ss'}\bar\delta_{a'b'}\right)\\
({\bar\Delta}_{ss'})_{a'a,-}&\equiv&0\\
({\bar\Delta}_{ss'})_{-,b'b}&\equiv&0\\
({\bar\Delta}_{ss'})_{-,-} &\equiv& \bar\delta_{ss'} \, .
\end{eqnarray}
\end{mathletters}

At this point, we can proceed with the iteration of the
quasiparticle expansion, and introduce the separable structure 
for the 4 subamplitudes of the system,
\begin{mathletters}
\begin{eqnarray}
(x_s)_{a'a,b'b}&=& |(s^{(2)})_{a'a} \rangle \tau^{(2)}_s
                           \langle (s^{(2)})_{b'b}|\\
(x_s)_{a'a,-}&=& |(s^{(2)})_{a'a}\rangle \tau^{(2)}_s
                           \langle (s^{(2)})_{-}|\\
({x^\dagger_s})_{-,b'b}&=&|(s^{(2)})_{-}\rangle \tau^{(2)}_s
                           \langle (s^{(2)})_{b'b}|\\
(x_s)_{-,-} &\equiv& |(s^{(2)})_{-}\rangle \tau^{(2)}_s
                           \langle (s^{(2)})_{-}| \,.
\end{eqnarray}
\label{supersep}
\end{mathletters}

\newpage

(As usual at this point, we must note that in case $s=0$ the
states $|s^{(2)}\rangle $ have no components in the no-pion sector).

In the upper side of Fig.~\ref{fig2} we represent two examples of
disconnected amplitude ${\bf x}_s$, both 
referring to the partition $s=1$, where the blue line
is not connected with the red and green ones. 
The box-like diagram on the left represents a process
connecting two states of the four-body sector. We have chosen the special 
case where the  ``in" and ``out" three-cluster states
coincide. In spite of this fact, the diagram does not represent
a {\it diagonal} matrix-element, because the three-cluster partition 
on the right coalesces into a  2+2 two-cluster partition,
while the same three-cluster partition on the left
has been originated from the break-up of the 3+1 partition.
The box-like diagram on the right represents
a disconnected production amplitude, where there is a collision
between nucleon ``2" and ``3" in presence of the nucleon ``1", with the
pion in the final three-cluster state. The 
selected production amplitude shows that the final three-cluster partition
derives from the break-up of the 3+1 two-cluster partition, however
it must be kept in mind that the final three-cluster state
can be obtained also from the 2+2 partition. This indicates that the 
role of the spectator nucleon (the blue line in the diagram)
is not passive at all, since it can still interact with the pion.
This contrasts with the standard three-particle case where the 
spectator merely plays a passive role.
In the lower part of the figure, the same amplitudes 
are represented in the form of quasiparticle diagrams, thus reproducing
Eqs.~(\ref{supersep}). The diagrams represent the processes
passing through the intermediate 
propagation of a multi-particle two-cluster state, where the 
nucleon ``1" is always disconnected from the other two. The pion, 
however, is shared between both parts without being physically  
exchanged from one to the other.

Introducing the new separable expansion of the subamplitudes in 
Eq.~(\ref{step1}), and folding the equation with the new states 
${\cal G}^{(3)}|s^{(2)}\rangle $ referring to the two-cluster partitions 
one obtains the final quasiparticle equation
\begin{equation}
X^{(2)}_{ss'}={Z^{(2)}_{ss'}}
+\sum_{s''} {Z^{(2)}_{ss''}}{{\cal G}^{(2)}_{s''}} {X^{(2)}_{s''s'}}
\label{step2}
\end{equation}
where 
\begin{equation}
{{\cal G}^{(2)}_{s}}= \tau^{(2)}_{s} \, 
\end{equation}
\begin{eqnarray}
Z^{(2)}_{ss'}= 
\langle s^{(2)}|{\cal G}^{(3)}\bar\Delta_{ss'}|{s'}^{(2)}\rangle
&\equiv& 
\langle (s^{(2)})_{-}| g_0 |({s'}^{(2)})_{-}\rangle\bar\delta_{ss'}+
\label{exchint}
%\\& & \nonumber
\sum_{a'(\subset s)}
\sum_{b'(\subset s')}
\sum_{a(\subset a',b')}
\langle (s^{(2)})_{a'a}| \tau_a |({s'}^{(2)})_{b'a}\rangle
(\bar\delta_{ss'}+\delta_{ss'}\bar\delta_{a'b'})
\end{eqnarray}
\begin{eqnarray}\nonumber
X^{(2)}_{ss'}&=& \langle s^{(2)}|{\cal G}^{(3)} 
                                  {{\bf X}_{ss'}}
                                   {\cal G}^{(3)} |{s'}^{(2)}\rangle  
\\ \nonumber
&\equiv& 
\langle (s^{(2)})_{-}| g_0 [ (X_{ss'})_{-,-} ] g_0  
 |({s'}^{(2)})_{-}\rangle 
%\\ \nonumber &&
+
\sum_{a'(\subset s)}
\sum_{b'(\subset s')}
\sum_{a(\subset a')}
\sum_{b(\subset b')}  
\langle (s^{(2)})_{a'a}| \tau_a [ (X_{ss'})_{a'a,b'b} ]\tau_b 
|({s'}^{(2)})_{b' b}\rangle  
\\ 
&+&
\sum_{a'(\subset s)}
\sum_{a(\subset a')} 
\langle (s^{(2)})_{a'a}| \tau_a  [ (X_{ss'})_{a'a,-} ] g_0
 |({s'}^{(2)})_{-}\rangle 
%\\ \nonumber &&
+ \sum_{b'(\subset s')}
  \sum_{b (\subset b')}  
+\langle (s^{(2)})_{-}| g_0 [ (X_{ss'})_{-,b'b} ] \tau_b 
 |(s'^{(2)})_{b' b}\rangle \, .
\end{eqnarray}

The expression Eq.~(\ref{step2}) represents the two-cluster connected-kernel
equation which solves the $\pi$NNN-NNN problem. It represents 
the translation within the quasiparticle formalism of the general
result represented by Eq.~(\ref{finalresult}). 
In spite of the fact that Eq.~(\ref{step2}) must be considered an 
approximated result holding only when the t-matrix separability 
is assumed, nevertheless the result should be considered under a
very general perspective because the representation of the 
t-matrix as a sum of separable terms is a mathematically 
converging procedure \cite{WEI} and approaches of this kind
have been demonstrated  to work numerically \cite{cghkpw} 
in few-body applications involving realistic nuclear interactions.

\begin{figure}[t]
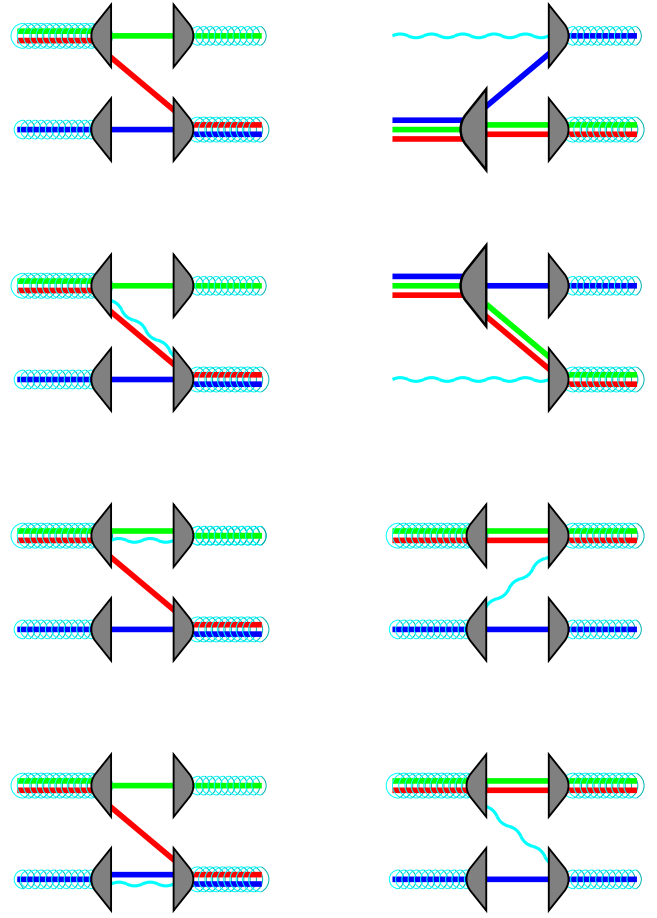

\begin{center}
\vskip 0 truecm
\Spicture{4.5in}{dfpd98th22diag3.ps}{}
\vskip -4.9 truecm
\end{center}
\caption[*]{Two-cluster exchange diagrams. The figure shows
the exchange diagrams contributing to the two-cluster potential
$Z^{(2)}_{ss'}$ of Eq.~(\ref{step2}).
The four diagrams on the left side contribute to $Z^{(2)}_{ss'}$
for $0\neq s \neq s'\neq 0$, while the two top diagrams on the right side
contribute for $0=s\neq s'$, and finally the remaining two bottom diagrams
contribute for $s=s'\neq 0$. There are no other diagrams to lowest order
(aside those obtained from permutation of the three colours)
and they are all connecting-type diagrams.\label{fig3}}
\leftline
\end{figure}

In Eq.~(\ref{step2}) the complete dynamics of Eq.~(\ref{finalresult}) is 
represented in terms of two-body multiparticle correlated states 
(bound-states, or resonances, etc., for the subsystems). 
They give a physically clear
description of the meaning of the general equations, 
otherwise difficult to interpret in terms  of processes or diagrams.
For instance, Eq.~(\ref{step2}) can be easily compared with the 
AGS quasiparticle equation for the standard three-particle problem:
Here, the equation is endowed with a fourth component
(the $s=0$ component) which does not appear in the AGS equations,
and the number of diagrams contributing to the Z-exchange terms
are considerably larger with some of them giving rise to 
totally new mechanisms. This can be seen in Fig.~\ref{fig3}
where the diagrams contributing to the Z terms [as expressed by
Eq.~(\ref{exchint})] are illustrated.

\section{bound state equation}
\label{boundsys}

This section is devoted to the discussion of the bound-state
equation for the $\pi$NNN-NNN system. The equation we derive
is in fact a bound-state equation for the three-nucleon system,
but has the special feature that it incorporates explicitly 
the pion dynamics (limited to the degree of freedom of one pion),
while in the standard approach this aspect is usually restricted
to the limits of the OPE tail of the NN interaction.

In the AGS approach, the three-nucleon 
bound state is associated to the homogeneous solution of the 
AGS equation. In close similarity, here we seek for the 
homogeneous solution of the AB equation for the $\pi$NNN system.

\leftlineup
\newpage

According to the matrix notation previously introduced,
we denote the homogeneous equation as 
\begin{equation}
|{\bf\Gamma}^{(3)}\rangle = {\bf V}^{(3)}{\bf G}_0^{(3)}| 
{\bf\Gamma}^{(3)}\rangle
\, , \label{eq:4.1}
\end{equation}
where $|{\bf \Gamma}^{(3)}\rangle$ represents the state eigenvector  
of the operator ${\bf V}^{(3)}{\bf G}_0^{(3)}$. Obviously
$|\Phi^{(3)}\rangle = {\bf G}_0^{(3)}|{\bf \Gamma}^{(3)}\rangle$ 
represents the analogous
eigenvector for the transposed kernel
\begin{equation}
|\Phi^{(3)}\rangle = {\bf G}_0^{(3)}{\bf V}^{(3)}| \Phi^{(3)}\rangle
\, .\label{eq:4.2}
\end{equation}
If we neglect all the couplings with the pion sector,
this last equation represents precisely the Schr\"odinger equation
for the three-nucleon system, with the constituents 
interacting  through pair-wise potentials and in such a case
$|\Phi^{(3)}\rangle$ denotes simply the complete three-body 
Schr\"odinger wavefunction. Once the one-pion degrees of freedom
are explicitly included into the theory, the equation acquires
the typical AB-like structure and couples the
three-nucleon Schr\"odinger wavefunction with the six Faddeev-like 
components referring to the partition of the $\pi$NNN system
into three clusters. Obviously, being the kernel of the homogeneous
equation the same as discussed in the previous sections, we have 
an equation whose kernel is not connected. 
%and we have again to face the problem
%of finding a connected-kernel equation.
%
We proceed as follows:

We introduce the partitions of the system into two clusters
and recall the interaction sumrule Eq.~(\ref{(III)}).
% write the potential sum
%\begin{equation}
%V^{(3)}=\left(
%\begin{array}{ll}
%V^{(3)}_{a,b} & V^{(3)}_{a,-} \\
%V^{(3)}_{-,b} & V^{(3)}_{-,-}
%\end{array}\right)
%=
%\left(
%\begin{array}{ll}
%\sum_{s=0,s} \sum_{a',b'\subset s} (v_s^{(3)})_{a'a,b'b} & 
%\sum_{s=0,s} \sum_{a'\subset s} (v_s^{(3)})_{a'a,-} \\
%\sum_{s=0,s} \sum_{b'\subset s} (v_s^{(3)})_{-,b'b} & 
%%\sum_{s=0,s} (v_s^{(3)})_{-,-}
%\end{array}\right)
%\end{equation}
Then, we
define the new two-cluster-partition components for the wavefunction:
\begin{mathletters}
\begin{eqnarray}
|(\Phi_s^{(2)})_{a'a}\rangle &=& 
G_0^{(3)} 
\sum_{b'(\subset s)}
\sum_{b(\subset b')}
 (v_s)_{a'a,b'b} 
|(\Phi^{(3)})_{b}\rangle \nonumber\\ &&+ 
G_0^{(3)} (v_s)_{a'a,-} 
|(\Phi^{(3)})_{-}\rangle \,,
\end{eqnarray}
and  
\begin{eqnarray}
|(\Phi_s^{(2)})_{-}\rangle &=& 
G_0^{(3)} 
\sum_{b'(\subset s)}
\sum_{b(\subset b')}
(v_s)_{-,b'b} 
|(\Phi^{(3)})_{b}\rangle \nonumber\\ &&+ 
G_0^{(3)} (v_s)_{-,-} 
|(\Phi^{(3)})_{-}\rangle \,, 
\end{eqnarray}
\end{mathletters}
where the first expression refers to components associated
to the 4-body sector
while the second one to the components in the three-nucleon space.

With this definition from the homogeneous 
equation for $|\Phi^{(3)}\rangle$, Eq.~(\ref{eq:4.2}), 
it is possible to express the
three-cluster components as sum over all the two-cluster partitions
\begin{mathletters}
\begin{eqnarray}
|(\Phi^{(3)})_{a}\rangle
= \sum_{s}\sum_{a'(\subset s)}  |(\Phi_s^{(2)})_{a'a}\rangle
\\
|(\Phi^{(3)})_{-}\rangle
= \sum_{s}  |(\Phi_s^{(2)})_{-}\rangle \, .
\end{eqnarray}
\end{mathletters}

From the last two equations it is possible 
to write a new homogeneous coupled equation
whose solution directly yields the components $|\Phi^{(2)}_s\rangle$.
We obtain
\newpage

\begin{mathletters}
\begin{eqnarray}
|(\Phi^{(2)}_s)_{a'a}\rangle &=&
\sum_{b'(\subset s)} 
\sum_{b(\subset b')}
G_0^{(3)}(v_s)_{a'a,b'b} 
\sum_{s'}
\sum_{c'(\subset s')}
|(\Phi^{(2)}_{s'})_{c'b}\rangle \nonumber \\
&+&
G_0^{(3)}(v_s)_{a'a,-} \sum_{s'}
|(\Phi^{(2)}_{s'})_{-}\rangle
\end{eqnarray}
and 
\begin{eqnarray}
|(\Phi^{(2)}_s)_{-}\rangle &=&
\sum_{b'(\subset s)}
\sum_{b(\subset b')}
G_0^{(3)}(v_s)_{-,b'b} \sum_{s'}\sum_{c'(\subset s')}
|(\Phi^{(2)}_{s'})_{c'b}\rangle \nonumber \\
&+&
G_0^{(3)}(v_s)_{-,-} \sum_{s'}
|(\Phi^{(2)}_{s'})_{-}\rangle
\end{eqnarray}
\end{mathletters}
\clearpage
\twocolumn[\hsize\textwidth\columnwidth\hsize\csname
@twocolumnfalse\endcsname
for the components in the 4-body and 3-nucleon sectors respectively.
With simple algebraic manipulations
we obtain
\begin{mathletters}
\begin{eqnarray}
\nonumber
\lefteqn{
|(\Phi^{(2)}_s)_{a'a}\rangle=
\sum_{b'(\subset s)} 
\sum_{c'(\subset s)} 
\sum_{b (\subset b',c')} 
G_0^{(3)}(v_s)_{a'a,b'b} 
|(\Phi^{(2)}_{s})_{c'b}\rangle
+
G_0^{(3)}(v_s)_{a'a,-} 
|(\Phi^{(2)}_{s})_{-}\rangle
} \\& &
+ 
\sum_{s'} \bar\delta_{ss'} 
\sum_{b'(\subset s)} 
\sum_{c'(\subset s')} 
\sum_{b (\subset b',c')} 
G_0^{(3)}(v_s)_{a'a,b'b} 
|(\Phi^{(2)}_{s'})_{c'b}\rangle
+
G_0^{(3)}(v_s)_{a'a,-} \sum_{s'} \bar\delta_{ss'}
|(\Phi^{(2)}_{s'})_{-}\rangle
\end{eqnarray}
and 
\begin{eqnarray}
\nonumber
\lefteqn{
|(\Phi^{(2)}_s)_{-}\rangle=
\sum_{b'(\subset s)} 
\sum_{c'(\subset s)} 
\sum_{b (\subset b',c')} 
G_0^{(3)}(v_s)_{-,b'b} 
|(\Phi^{(2)}_{s})_{c'b}\rangle
+
G_0^{(3)}(v_s)_{-,-} 
|(\Phi^{(2)}_{s})_{-}\rangle 
} \\& &
+ \sum_{s'} \bar\delta_{ss'} 
\sum_{b'(\subset s)} 
\sum_{c'(\subset s)} 
\sum_{b (\subset b',c')} 
G_0^{(3)}(v_s)_{-,b'b} 
|(\Phi^{(2)}_{s'})_{c'b}\rangle
+
G_0^{(3)}(v_s)_{-,-} \sum_{s'} \bar\delta_{ss'}
|(\Phi^{(2)}_{s'})_{-}\rangle
\end{eqnarray}
\end{mathletters}
The last two equations can be rewritten as
\begin{mathletters}
\begin{eqnarray}
\nonumber
\lefteqn{
|(\Phi^{(2)}_s)_{a'a}\rangle  - 
\sum_{b'(\subset s)} 
\sum_{b (\subset b')} 
G_0^{(3)} (v_s)_{a'a,b'b} 
|(\Phi^{(2)}_s)_{b'b}\rangle
-G_0^{(3)} (v_s)_{a'a,-} 
|(\Phi^{(2)}_s)_{-}\rangle 
}\\& &
 =
\sum_{s'}
\sum_{b'(\subset s)} 
\sum_{c'(\subset s')} 
\sum_{b (\subset b',c')}  
G_0^{(3)}(v_s)_{a'a,b'b} \left(\bar\delta_{ss'}+
\delta_{ss'}\bar\delta_{b'c'}\right)
|(\Phi^{(2)}_{s'})_{c'b}\rangle
+
\sum_{s'}
G_0^{(3)}(v_s)_{a'a,-} \bar\delta_{ss'} 
|(\Phi^{(2)}_{s'})_{-}\rangle 
\end{eqnarray}
and 
\begin{eqnarray}
\nonumber
\lefteqn{
|(\Phi^{(2)}_s)_{-}\rangle  - 
\sum_{b'b} G_0^{(3)} (v_s)_{-,b'b} 
|(\Phi^{(2)}_s)_{b'b}\rangle
-G_0^{(3)} (v_s)_{-,-} 
|(\Phi^{(2)}_s)_{-}\rangle 
} \\& &
 =
\sum_{s'}
\sum_{b'(\subset s)} 
\sum_{c'(\subset s')} 
\sum_{b (\subset b',c')}  
G_0^{(3)}(v_s)_{-,b'b} \left(\bar\delta_{ss'}+
\delta_{ss'}\bar\delta_{b'c'}\right)
|(\Phi^{(2)}_{s'})_{c'b}\rangle
+
\sum_{s'}
G_0^{(3)}(v_s)_{-,-} \bar\delta_{ss'} 
|(\Phi^{(2)}_{s'})_{-}\rangle . 
\end{eqnarray}
\end{mathletters}

From these, employing the equations for the subsystem amplitudes,
Eqs.~(\ref{subdynamics}), 
it is possible to obtain the final bound-state equation,
\begin{mathletters}
\begin{eqnarray}
\nonumber
\lefteqn{
|(\Phi^{(2)}_s)_{a'a}\rangle 
 =
} \\& &
\sum_{s'}
\sum_{b'(\subset s)} 
\sum_{c'(\subset s')} 
\sum_{b (\subset b',c')}  
G_0t_aG_0(t_s)_{a'a,b'b} \left(\bar\delta_{ss'}+
\delta_{ss'}\bar\delta_{b'c'}\right)
|(\Phi^{(2)}_{s'})_{c'b}\rangle 
+
\sum_{s'}
G_0t_aG_0(t_s)_{a'a,-} \bar\delta_{ss'} 
|(\Phi^{(2)}_{s'})_{-}\rangle 
\end{eqnarray}
and 
\begin{eqnarray}
\nonumber
\lefteqn{
|(\Phi^{(2)}_s)_{-}\rangle  =
} \\& &
\sum_{s'}
\sum_{b'(\subset s)} 
\sum_{c'(\subset s')} 
\sum_{b (\subset b',c')} 
g_0(t_s)_{-,b'b} \left(\bar\delta_{ss'}+
\delta_{ss'}\bar\delta_{b'c'}\right)
|(\Phi^{(2)}_{s'})_{c'b}\rangle
+
\sum_{s'}
g_0(t_s)_{-,-} \bar\delta_{ss'} 
|(\Phi^{(2)}_{s'})_{-}\rangle \, . 
\end{eqnarray}
\label{boundst}
\end{mathletters}
\leftline
]
Eqs.~(\ref{boundst}) represent the generalization of the
bound-state three-nucleon equation and include
in the three-nucleon dynamics also the pion dynamics.
The bound-state wavefunction corresponds to
the solution of the homogeneous equation whose kernel is
transposed with respect to that of Eq.~(\ref{finalresult})
for the scattering amplitudes.

In the no-pion sector the complete three-nucleon wavefunction
is given simply by the sum over the three components $s$=1,2,3
(the $s=0$ case has no direct component in the no-pion sector):
\begin{equation}
|(\Phi^{(3)})_{-}\rangle=
\sum_{s}
|(\Phi^{(2)}_{s})_{-}\rangle  \,.
\label{3Ncomponent}
\end{equation}
This result is similar to that obtained in standard Faddeev theory,
where the three-nucleon bound state is given by the sum over the 
three Faddeev components.

One may consider at this point the other component of the wavefunction,
the one acting in the four-body sector (obviously, 
in the standard three-nucleon theory
these components are set identically to zero).
The wavefunction $|\Phi^{(3)}\rangle$
in the four-body sector spans the six three-cluster partitions
of the $\pi$NNN system. In a standard 4-body theory the complete 
wavefunction is given by the sum over these six Faddeev 
components. In the present theory we have to take into 
account the fact that a contribution to the wavefunction
may arise by pion \clearpage
\twocolumn[\hsize\textwidth\columnwidth\hsize\csname
@twocolumnfalse\endcsname
emission from the pure three-nucleon component,
therefore the 4-body component to the three-nucleon bound-state
wavefunction is given by
\begin{equation}
|\Phi^{(4)}\rangle = \sum_{a=1}^6
|\Phi^{(3)}_{a}\rangle
+G_0(\sum_{i=1}^3 f_i) |\Phi^{(3)}_{-}\rangle 
=\sum_s
\sum_{a'(\subset s)}
\sum_{a(\subset a')}
|(\Phi_s^{(2)})_{a'a}\rangle
+G_0(\sum_{i=1}^3 f_i) \sum_s |(\Phi^{(2)}_s)_{-}\rangle 
\, ,
\label{4Bcomponent}
\end{equation}
\leftline
]
where the second addend comes from the coupling with the  
pure the three-nucleon space. 
%This equation is the counterpart in the language of the wavefunction
%of the analogous equation (\ref{t44}) for the T-matrix.

\section{rearrangement and break-up amplitudes}
\label{amplitudes}

In Sect.~\ref{clusdec} we have restricted the discussion to the fully unclusterized
amplitudes (four-to-four) or at most to the three-to-three amplitudes.
Then in Sect.~\ref{quasiparticle} we have given the rules to calculate
$T(1|1)$ with the quasiparticle formalism.
It is clear that from the phenomenological point of view the most interesting 
amplitudes are between channels involving the 
two-cluster partitions, or amplitudes
where at least the incoming state refers to an asymptotic configuration
where the system is partitioned into two clusters.
To obtain such amplitudes, we start from the three-to-three
amplitudes and apply the residue method. To this end we introduce the 
homogeneous equations associated with the two-cluster partition:
\begin{equation}
|{\gamma^{(2)}_{s}}(E_s)\rangle 
= {\bf v}_{s}(E_s){\bf G}_0^{(3)}(E_s)|{\gamma^{(2)}_{s}}(E_s)\rangle \, ,
\label{gamma_s}
\end{equation}
where for each $s\ne$0 the state $|{\gamma^{(2)}_{s}}\rangle$
represents a channel vector with one component in the no-pion sector
(this substitute the Faddeev component of the standard 3N theory)
and five components in the one-pion sector (corresponding to all possible 
chains of partitions starting from the 2+2 and 3+1 partitions
compatible with $s$). For the special case $s=0$, the same equation
couples only the three chains of partitions which start
from the $\pi$+ (NNN) separation in two cluster, and have
no components in the 3N sector. Similarly, one can introduce also the
corresponding homogeneous equation for the {\it bra} states
\begin{equation}
\langle {\gamma^{(2)}_{s}}(E_s)| 
=
\langle {\gamma^{(2)}_{s}}(E_s)| 
{\bf G}_0^{(3)}(E_s)
{\bf v}_{s}(E_s)
\label{gamma_s_bra} \, .
\end{equation}

Obviously for each $s$, with the transforming relations
\begin{eqnarray}
|{\gamma^{(2)}_{s}}\rangle &= {\bf v}_{s} |{\phi^{(2)}_{s}}\rangle \, , \\
 |{\phi^{(2)}_{s}}\rangle  &= {\bf G}_0^{(3)} |{\gamma^{(2)}_{s}}\rangle \, ,
\end{eqnarray}
it is 
possible to associate an asymptotic channel state
statisfing a bound-state-type equation (the energy dependence
has been omitted) 
for the two noninteracting fragments 
\begin{equation}
|{\phi^{(2)}_{s}}\rangle = {\bf G}_0^{(3)}{\bf v}_{s} |{\phi^{(2)}_{s}}
\rangle \, .
\end{equation}
We can view explicitly how in case $s\ne$0 the new equation
couples the chain-space in the four-body sector with the 
Faddeev components of the 3N space by writing in detail
the homogeneous equation
\begin{mathletters}
\begin{eqnarray}
|(\phi^{(2)}_{s})_{a'a}\rangle &=& \sum_{b(\subset a')}
G_0t_a \bar\delta_{ab} |(\phi^{(2)}_{s})_{a'b}\rangle 
\nonumber\\ &&
+G_0t_aG_0(f_s)_{a'a} |(\phi^{(2)}_{s})_{-}\rangle \\
|(\phi^{(2)}_{s})_{-}\rangle &=& 
\sum_{b'(\subset s)} 
\sum_{b (\subset b')} 
g_0(f_{s}^{\dagger})_{b'b}|(\phi^{(2)}_{s})_{b'b}\rangle 
\nonumber\\ &&
+g_0 {\cal V}_{s} |(\phi^{(2)}_{s})_{-}\rangle \, ,
\end{eqnarray}
\end{mathletters}
while for $s=0$ we have a standard three-component (Faddeev-like)
3N bound-state equation, with the pion acting as a spectator.
In case the couplings between the 
two spaces are switched off, each coupled six-component equation
for $s\ne$0 decouple into the three different equations.
One single component homogeneous equation for the NN  
pair in presence of a spectator nucleon plus one three-component Faddeev equation for the 3+1 partition and one 
analogous two-component coupled equation for the 
corresponding 2+2 partition. With the meson-nucleon vertex interaction 
turned on, these three different equations merge in one single
coupled equation.

At energies $E_s$ corresponding to nontrivial solutions
of the homogeneous equations it follows 
that the solution of the inhomogeneous equation
${\bf t}_s$ has a pole, and around such values
the t-matrix for the subsystem can be represented
in polar form
\begin{equation}
{\bf t}_{s}(z)\simeq |\gamma_{s}^{(2)}\rangle{1\over z-E_s}
\langle\gamma_{s}^{(2)}|+ ...
\end{equation}
where the omitted contributions are nonsingular background remainders.

According to the residue method, the clusterized transition
amplitudes can be obtained from the general expression for 
${\bf T}^{(3)}$, Eqs.~(\ref{(IV)}),
 by extracting  the residues once the poles
of the subamplitudes are exhibited.
For instance, if we assume $s=$0 and $s'\ne$0
and assuming that for $E_s$ and $E_{s'}$ the associated 
homogeneous equations have a nontrivial (bound-state or narrow resonance) 
solution, then the corresponding two-cluster transition amplitude 
emerges as the residue of the double singularity in ${\bf T}^{(3)}$
\begin{equation}
{\bf T}^{(3)}=|{\bar \gamma}_{s'}^{(2)}\rangle {{\cal T}_{s's}
\over (z-E_{s'})(z-E_{s})}
\langle{\bar \gamma}_{s}^{(2)}| + ... \, \, ,
\end{equation}
where the state vectors $|{ \gamma}_s^{(2)}\rangle$  have been contracted
by summing over the the two-body partitions
of the four-body sector,  $|({\bar \gamma}_s^{(2)})_a\rangle
=\sum_{a'(\subset s)} |({ \gamma}_s^{(2)})_{a'a}\rangle$. 
\clearpage

\twocolumn[\hsize\textwidth\columnwidth\hsize\csname
@twocolumnfalse\endcsname

The two-cluster transition matrix element is given by 
\begin{eqnarray}\nonumber
\lefteqn{
{\cal T}_{s's}
= \langle{\phi^{(2)}_{s'}}|U_{s's}|{\phi^{(2)}_{s}}\rangle
} \\ &=& 
%\sum_{a\subset a'\subset s}
%\sum_{b\subset b'\subset s'}
\sum_{a'(\subset s')}  
\sum_{a (\subset a')} 
\sum_{b'(\subset s)}  
\sum_{b (\subset b')} 
\langle(\phi^{(2)}_{s'})_{a'a}|(U_{s's})_{a'a,b'b}
|(\phi^{(2)}_{s})_{b'b}\rangle %\nonumber \\& &
+
%\sum_{a\subset a'\subset s}
\sum_{b'(\subset s)}  
\sum_{b (\subset b')} 
\langle(\phi^{(2)}_{s'})_{-}|(U_{s's})_{-,b'b}
|(\phi^{(2)}_{s})_{b'b}\rangle \, ,
\end{eqnarray}
where in the last expression on the right 
the components for $s'$ acting in each sector of the theory
have been explicitly given.
In this approach, such an amplitude represents
the process $\pi$ + (NNN)$\rightarrow$ N + (NN) where
the contributions of the type ($\pi$N) + (NN), and  
N + (NN$\pi$) are both dynamically included together with the
N +(NN) partition.

We may at this point report on break-up reaction amplitudes, 
such as $\{$ $\pi$ (NNN) $\rightarrow$ N N N $\}$,  
$\{$ $\pi$ (NNN) 
$\rightarrow$ $\pi$ N (NN)$\}$ and
finally $\{$ $\pi$ (NNN) $\rightarrow$ N N N $\pi$ $\}$.
The first two can be obtained from ${\bf T}^{(3)}$ by 
extraction of the residue of a single two-cluster partition 
(bound-state) singularity, while
for the last case one has to consider the single residue
from Eq.~(\ref{lunga}).
We have (with $s$=0)
\begin{eqnarray} \nonumber
%\lefteqn{
{\cal T}_{0s}(NNN\leftarrow\pi(NNN))
%= \sum_{s'\ne 0} \langle\chi^{(3)}_{0}
%|u^{(3)}_{s'}G_0^{(3)}U_{s's}|\Phi^{(3)}_{s}\rangle
%} \\ %\nonumber
&=&
%\sum_{a\subset a'\subset s'}
%\sum_{b\subset b'\subset s}
\sum_{s'}
\sum_{a'(\subset s')}  
\sum_{a (\subset a')} 
\sum_{b'(\subset s)}  
\sum_{b (\subset b')} 
\langle(\chi^{(3)}_0)|(t_{s'})_{-,a'a}G_0t_aG_0(U_{s's})_{a'a,b'b}
|(\phi^{(2)}_{s})_{b'b}\rangle
\\&&
+\sum_{s'}
%\sum_{b\subset b'\subset s}
\sum_{b'(\subset s)}  
\sum_{b (\subset b')} 
\langle(\chi^{(3)}_0)|(t_{s'})_{-,-}g_0(U_{s's})_{-,b'b}
|(\phi^{(2)}_{s})_{b'b}\rangle\,\, ,%\\ \nonumber 
\end{eqnarray}
%and
\begin{eqnarray} \nonumber
%\lefteqn{
{\cal T}_{as}((NN)\pi N \leftarrow\pi(NNN))
%= \langle{\bf\phi^{(2)}}_{s}
%\sum_{s'\ne 0}|t_{s'}G_0^{(3)}U_{s's}|{\bf\phi^{(2)}}_{s}\rangle
%} \\  
&=&
\sum_{s'}
\sum_{a'(\subset s')}
\sum_{c'(\subset s')}  
\sum_{c (\subset c')} 
\sum_{b'(\subset s)}  
\sum_{b (\subset b')} 
%\sum_{b\subset b'\subset s}\sum_{c\subset c'\subset s'}
\langle \phi^{(3)}_a|(t_{s'})_{a'a,c'c}G_0t_cG_0(U_{s's})_{c'c,b'b}
|(\phi^{(2)}_{s})_{b'b}\rangle
\\ & &
+
\sum_{s'}
\sum_{a'(\subset s')}
\sum_{b'(\subset s)}  
\sum_{b (\subset b')} 
%\sum_{a'\subset s'}\sum_{b\subset b'\subset s}
\langle \phi^{(3)}_a|(t_{s'})_{a'a,-}g_0(U_{s's})_{-,b'b}
|(\phi^{(2)}_{s})_{b'b}\rangle \,\, .
\end{eqnarray}

In this last case $a$ represents the selected NN pair,
in presence of two remaining spectator particles,
and $\langle \phi_a^{(3)}(E_a)|=
\langle \phi_a^{(3)}(E_a)|v_aG_0(E_a)$ is the asymptotic 
three-cluster channel with two bound nucleons in presence of two
spectator particles.
The amplitude referring to the process with
four outgoing fragments is
\begin{eqnarray}
\nonumber 
%\lefteqn{
{\cal T}_{0s}( \pi NNN\leftarrow\pi(NNN))
%} {\ \ \ } 
%& &\\ \nonumber 
&=\ &
\sum_{s'}
\sum_{a'(\subset s')}
\sum_{a (\subset a')}
\sum_{c'(\subset s')}  
\sum_{c (\subset c')} 
\sum_{b'(\subset s)}  
\sum_{b (\subset b')} 
%\sum_{a\subset a'\subset s'}
%\sum_{b\subset b'\subset s'}
%\sum_{c\subset c'\subset s}
\langle\chi^{(4)}_{0}|
t_a G_0 (t_{s'})_{a'a,c'c}G_0t_cG_0
(U_{s's})_{c'c,b'b}|(\phi^{(2)}_{s})_{b'b}\rangle
 \\ \nonumber & &
+
\sum_{s'}
\sum_{a'(\subset s')}
\sum_{a (\subset a')}
\sum_{b'(\subset s)}  
\sum_{b (\subset b')} 
%\sum_{a\subset a'\subset s'}
%\sum_{b\subset b'\subset s'}
\langle\chi^{(4)}_{0}|
t_a G_0 (t_{s'})_{a'a,-}g_0
(U_{s's})_{-,b'b}|(\phi^{(2)}_{s})_{b'b}\rangle
\\ \nonumber & &+
\sum_{s'}
\sum_{c'(\subset s')} 
\sum_{c (\subset c')} 
\sum_{b'(\subset s)}  
\sum_{b (\subset b')} 
%\sum_{b\subset b'\subset s'}
%\sum_{c\subset c'\subset s}
\langle\chi^{(4)}_{0}|
(\sum_{i}f_i) g_0 (t_{s'})_{-,c'c}G_0t_cG_0
(U_{s's})_{c'c,b'b}|(\phi^{(2)}_{s})_{b'b}\rangle
\\& &
+
\sum_{s'}
\sum_{b'(\subset s)}  
\sum_{b (\subset b')} 
%\sum_{b\subset b'\subset s'}
\langle\chi^{(4)}_{0}|
(\sum_{i}f_i) g_0 (t_{s'})_{-,-}g_0
(U_{s's})_{-,b'b}|(\phi^{(2)}_{s})_{b'b}\rangle \, .
\end{eqnarray}
\leftline
%Clearly, following the same formalism
%one could write down explicitly the remaining amplitudes;
%however they do not introduce any novelty with respect to the 
%ones already given and therefore will be omitted.
]
The states $\langle \chi^{(3,4)}_{0}|$ represent, respectively,
the free three-nucleon and the four-particle asymptotic  waves.

It is worthwhile to comment one aspect common to
all these amplitudes; namely in all the physical reaction 
processes one has to sum over the possible (for a given $s$) 
two-cluster partitions of the four-body sector (herein denoted 
with $a'$, $b'$ and $c'$).
The $s$=0 partition of the system is an exception only because 
it contains just one on these partitions (Tab.~\ref{tab3}).
In other words, while the indices $a'$ etc. are fundamental
in the determination of the dynamical equation,
they do not appear in the physical amplitudes, since these last
quantities have to refer only to partitions of the whole system.
This specific aspect of the coupled $\pi$NNN
theory emerges from the structure of the dynamical equations,
which are labelled 
in the four-body sector
by chains of partitions, 
{\em e.g.}, the pair $a'a$ with $a\subset a'$, 
while the physical amplitudes refer only to physical
partitions of the complete system into clusters.

In the remaining part of this section
we show how the quasiparticle method 
can be extended to the present situation
to calculate the clusterized amplitudes.
Previously, starting from the separable expansion 
$
%\begin{equation}
t_a^{(3)}=|a^{(3)}\rangle \tau_a^{(3)} \langle a^{(3)}|
%\end{equation}
$
we arrived at the equation
$
%\begin{equation}
{\bf x}_s = {\bf z}_s + {\bf z}_s {\cal G}^{(3)} {\bf x}_s
%\end{equation}
$
where we observed that the operators ${\bf z}_s$ and ${\bf x}_s$
act in the chain-of-partition space of the four-body
sector, and (for $s\ne$0) in the space of two-cluster
partitions of the three-nucleon sector.
A second iteration of the quasiparticle method consisted in the
exhibition of the separable structure for ${\bf x}_s$,
$
%\begin{equation}
{\bf x}_s=|s^{(2)}\rangle \tau_s^{(2)} \langle s^{(2)}|
%\end{equation}
$
which allowed to derive the equation
\begin{equation}
X_{ss'}^{(2)} = Z_{ss'}^{(2)} + 
\sum_{s''} Z_{ss''}^{(2)} {\cal G}_{s''}^{(2)} X_{s''s'}^{(2)}.
\end{equation}

In the general case, we have seen in this Section 
that the two-cluster rearrangement
amplitudes
can be written as
\begin{equation}
{\cal T}_{ss'}=\langle{\bf \phi}_s^{(2)}| U_{ss'}|{\bf\phi}_s^{(2)}\rangle
=\langle{\bf \gamma}_s^{(2)}|G_0^{(3)} U_{ss'}G_0^{(3)}
| {\bf\gamma}_{s'}^{(2)}\rangle \, .
\end{equation}
 At this point, taking advantage of the 
separable expression for the t-matrix in $(G^{(3)})_a = G_0 t_a G_0$
we obtain
 \begin{equation}
 {\cal T}_{ss'}=
\langle {\cal s}_s^{(2)}|{\cal G}^{(3)} {\bf X}_{ss'}{\cal G}^{(3)}| 
{\cal s}_{s'}^{(2)}\rangle
\end{equation}
where 
\begin{equation}
|{\cal s}_s^{(2)}\rangle
\equiv \left(
\begin{array}{c}
| ({\cal s}_s^{(2)})_{a'a}\rangle \\
| ({\cal s}_{s}^{(2)})_{-}\rangle
\end{array}\right)
=\left( 
\begin{array}{c}
\langle a | G_0 
|(\gamma_s^{(2)})_{a'a}\rangle
\\
|(\gamma_s^{(2)})_{-}\rangle
\end{array}\right) \, .
\end{equation}

Stated in this form, this implies that to calculate the reaction
amplitude ${\cal T}_{ss'}$ we have to solve the equation for ${\bf X}_{ss'}$ 
and the homogeneous equations for $|{\bf \gamma}^{(2)}_s\rangle$ to produce the states
$|{\cal s}^{(2)}_s\rangle$. 
It is however possible to exploit the quasiparticle/separable structure 
for the t-matrix in the homogeneous equations for the 
$|{\bf \gamma}_s^{(2)}\rangle$
and transform it into an homogeneous equation for the
states $|{\cal s}_s^{(2)}\rangle$.
This can be done by writing explicitly Eq.~(\ref{gamma_s}),
folding its components in the four-body space 
with $\langle a|G_0$ to the left, and using
Eq.~(\ref{rank-one}). One obtains
the homogeneous equation for the subsystem dynamics in 
quasiparticle form
\begin{eqnarray}
|{\cal s}^{(2)}_s(E_s)\rangle = {\bf z}_s(E_s) {\cal G}^{(3)}(E_s)
|{\cal s}^{(2)}_s(E_s)\rangle
\, .
\end{eqnarray}

The corresponding inhomogeneous version of this equation has been 
already given in Eq.~(\ref{x_s}), where the block matrices
${\bf z}_s$ and ${\cal G}^{(3)}$ have been explicitly given. 
The fact that the states $|{\cal s}^{(2)}_s\rangle$ are eigensolution
of the kernel for ${\bf x}_s$ implies that
a particularly convenient expression arises when these
states are used as a basis for the quasiparticle 
expansion of ${\bf x}_s$, {\it i.e.}
$| s^{(2)} \rangle
\equiv 
|{\cal s}_s^{(2)} (E_s) \rangle $. 
In that case the pole structure of ${\bf x}_s$
for $z\sim E_s$  naturally emerges
in the quasiparticle expansion, 
\begin{equation}
{\bf x}_s\simeq |{s}^{(2)}\rangle {1\over z-E_s}\langle
{s}^{(2)}| \, .
\end{equation}
Treatments of the like, based upon the idea of pole dominance
of the three-body subsystem operators in the kernel of  the standard 
four-body equations, have been suggested in various forms \cite{smf,adikowa}
(for a short review on recent applications, see also Ref.~\cite{ssehs}).

With the idea of pole dominance, 
the solution of the final Eq.~(\ref{step2}),
{\it i.e.} the amplitudes $X^{(2)}_{ss'}$, when calculated on-shell, 
directly yield  the two-cluster reaction amplitudes ${\cal T}_{ss'}$,
 \begin{eqnarray}\nonumber
X^{(2)}_{ss'}&=&\langle s^{(2)}|{\cal G}^{(3)} {\bf X}_{ss'} {\cal G}^{(3)}
|{s'}^{(2)}\rangle \\
&=&
\langle{\bf\gamma}_s^{(2)}|
G_0^{(3)} U_{ss'}G_0^{(3)}| 
{\bf\gamma}_{s'}^{(2)}\rangle
={\cal T}_{ss'} \, .
\end{eqnarray}

\section{summary, conclusions and outlook}
\label{conclusions}

This paper deals with the formulation of the three-nucleon problem
with inclusion of an additional pionic degree of freedom. The subject
implies to outrun the rather difficult question of developing
a few-body integral-equation approach with particle-nonconserving 
interactions.
The problem is solved in the truncated Hilbert space defined
by states with at most one pion, {\it e.g.} the coupled $\pi$NNN-NNN
space. Attempts in this direction have been made before, 
but the solution here developed is original and more complete.

The first, crucial step has been the clarification of
a rather delicate question of fragmentation
of the system into two clusters (Tab.~\ref{tab3}). 
The meson-nucleon vertex
interaction radically changes the cluster properties
of the system with respect to the standard case. For instance,
in the standard four-body case, the 3+1 and 2+2 partitions
are not coupled, while in the $\pi$NNN system three over four
3+1 partitions are coupled to the corresponding 
2+2 partitions, with the 2+1 partition of the three-nucleon 
space acting as a doorway state. Only the remaining forth
3+1 partition (the one with the spectator pion) keeps its 
standard 4-body role and conserves the number of particles. 
In other words, the two-cluster partitions of the coupled $\pi$NNN-NNN
system are transversal with respect to the number of particles,
since three partitions do not conserve the number of particles
while the fourth (denoted $s=0$) does so.

Then, the solution of the problem has been obtained
by rewording the multiparticle collision theory 
of Grassberger-Sandhas in terms tools:
%\begin{itemize}
%\item
{(i)} 
A disconnected dynamical equation of LS-type for multi-cluster processes
of the whole system ($T=V+VG_0T$).
%\item
{(ii)} A similar integral-equation approach for the subsystem dynamics,
which allows the systematic classification of the disconnected diagrams.
($t_s=v_s+v_sG_0t_s$).
%\item
{(iii)} A sum-rule equation for the multi-cluster interaction, which 
prevents the overcounting or undercounting
($V=\sum_s v_s$).
%\item
{(iv)} The systematic extraction of the disconnected contributions
from the initial multi-cluster collision amplitude of the whole system.
($T=\sum_s t_s + \sum_{ss'} t_s G_0 U_{ss'} G_0 t_{s'}$).
%\end{itemize}
In N-body scattering theory, from (i-iv), it is possible
to obtain a new dynamical equation (of AGS type) 
for the amplitudes $U_{ss'}$, which can be formally recast into the LS form. 
Hence, by repeated applications of the method, it is possible to extract 
gradually all the disconnected subamplitudes, thereby obtaining at the end a 
connected-kernel formulation of the quantum N-body problem.

In the present paper we have shown that these multiparticle tools work also
in presence of particle-non-conserving interactions, at least
if we choose
as starting point the integral-equation  approach of
Thomas-Rinat-Afnan-Blankleider-Avishai-Mizutani. Obviously,
the multiparticle method here developed cannot heal the typical
limitations of such input formalisms 
based on truncation of the Hilbert spaces. 
%But it has {\it not} been demonstrated either 
%that these multiparticle tools do not work when 
%applied to field-theoretic approaches
%which overcome such limitations.

The final formulation is represented by the set of Eqs.~(\ref{finalresult}),
which generalizes the AGS three-nucleon approach.
We have discussed this result in the light of the quasiparticle 
formalism, which allows a physically more transparent interpretation
in terms of coalescence diagrams. Within this formalism the OPE
diagram between two nucleons is treated at the same level of 
the particle-exchange diagrams between multiparticle clusters (Fig.~\ref{fig1}). 
The final equation, Eq.~(\ref{step2}), represents an effective
two-cluster equation, and the corresponding effective multichannel
potential is given exclusively by connected-type particle-exchange 
diagrams (Fig.~\ref{fig3}).
In the same framework, we have also
given the rules to calculate the various multiparticle
collision processes, including rearrangement reactions, break-ups,
pion-induced absorptions and productions.  

Finally, we have formulated the bound-state problem, Eq.~(\ref{boundst}).
The equation incorporates the dynamical effect of one pion in
the three-nucleon bound-state equation. From the solution of the
equation it is possible to calculate the bound-state wavefunction
in both its NNN and $\pi$NNN components, through Eqs.~(\ref{3Ncomponent})
and (\ref{4Bcomponent}), respectively. This approach represents
a formulation of the three-nucleon problem going beyond
a description in terms of pure two-nucleon potentials, which is 
notoriously inadequate (as shown in Ref.~\cite{HWKNG} and in the references 
therein contained). 
It does not require, on the other hand, the employment of  
three-nucleon forces (3NF) and the associated additional
fixing of new parameters (typically, 3NF cutoffs);
3NF represent an approximate, effective way to describe the underlying
meson dynamics in the three-nucleon system and the separation
between 2NF and 3NF requires an high level of consistency.
In Eq.~(\ref{boundst}), {\it all} possible combinations
of 3N diagrams reducible  to two-particle interactions ($\pi$N or NN)
while the dynamical pion is ``in flight" are taken into account
through the couplings with the four-body sector:
these contributions obviously represent a fraction
of the three-nucleon force, presumably the part with the 
longest range.

Of course, it is always possible to consider in principle
the additional effect of a residual 3NF, representing more complex 
(and shorter range) diagrams with at least two dynamical pions in the 
intermediate states, or conversely 
to attempt the more ambitious program of extending the present approach
to include multi-pion degrees of freedom. That would also reduce the 
effect of the main limitation implied by the approach, wherein 
the input interactions
have to be extracted from the disconnected $\pi$NNN-NNN amplitudes,
rather than from the $\pi$N subsystem amplitude.

%\begin{table}

%\begin{center}

%\begin{tabular}{|cc|}
%\hline
%$a$ & Partitions \\
%\hline
%1 & $(\pi\ N_1)\ N_2\ N_3 $\\
%2 & $(\pi\ N_2)\ N_3\ N_1 $\\
%3 & $(\pi\ N_3)\ N_1\ N_2 $\\
%4 & $(N_1\ N_2)\ N_3\ \pi $\\
%5 & $(N_2\ N_3)\ N_1\ \pi $\\
%6 & $(N_3\ N_1)\ N_2\ \pi $\\
%- & $ N_1\ N_2\ N_3 $\\
%\hline
%\end{tabular}

%\end{center}

%\caption[*]{The seven three-cluster partitions of the $\pi$NNN-NNN system.}
%\label{tab1}
%\end{table}

\end{document}